\documentclass[twocolumn,showpacs,preprintnumbers,amsmath,amssymb,pra,aps,10pt]{revtex4-1}

\usepackage{graphicx}
\usepackage{dcolumn}
\usepackage{bm}
\usepackage[T1]{fontenc}
\newcommand{\etal}{\textit{et al$\ $}}

\begin{document}
\preprint{Submitted to: JOURNAL OF PHYSICS: CONDENSED MATTER}
\title{Phase separation in a lattice model of a superconductor with pair hopping}
\author{Konrad Kapcia}%
    \email[corresponding author; e-mail: ]{kakonrad@amu.edu.pl}
\affiliation{Faculty of Physics, Adam Mickiewicz University, Umultowska 85, 61-614 Pozna\'n, Poland}
\author{Stanis\l{}aw Robaszkiewicz}%
\affiliation{Faculty of Physics, Adam Mickiewicz University, Umultowska 85, 61-614 Pozna\'n, Poland}
\author{Roman Micnas}%
\affiliation{Faculty of Physics, Adam Mickiewicz University, Umultowska 85, 61-614 Pozna\'n, Poland}
\date{March 3, 2012}
\begin{abstract}
We have studied the extended Hubbard model with pair hopping in the atomic limit for arbitrary electron density and chemical potential. The
Hamiltonian considered consists of (i)~the effective on-site interaction $U$ and (ii)~the intersite charge exchange interactions $I$, determining the hopping of electron pairs between nearest-neighbour sites. The model can be treated as a simple effective model of a~superconductor with very short coherence length in which electrons are localized and only electron pairs have possibility of transferring.
The phase diagrams and thermodynamic properties of this model have been determined within the variational approach, which treats the on-site interaction term exactly and the intersite interactions within the mean-field approximation. We have also obtained rigorous results for a~linear chain (\mbox{$d=1$}) in the ground state. Moreover, at \mbox{$T=0$} some results derived within the random phase approximation
(and  the spin-wave approximation)
for \mbox{$d=2$} and \mbox{$d=3$} lattices  and within the low density expansions for \mbox{$d=3$} lattices are presented.
Our investigation of the general case (as a function of the electron concentration $n$ and as a function of the chemical potential $\mu$) shows that, depending on the values of interaction parameters, the system can exhibit not only the homogeneous phases: superconducting (SS) and nonordered (NO), but also the phase separated states (PS: \mbox{SS--NO}).
The system considered exhibits interesting multicritical behaviour including tricritical points.
\end{abstract}

\maketitle


\section{Introduction}\label{sec:intro}

There has been much interest in superconductivity with very short coherence length. This interest is due to its possible relevance to high temperature superconductors (cuprates, doped bismuthates, iron-based systems, fullerenes) and also to the several other exotic superconducting materials  (for a review, see \cite{MRR1990,AAS2010} and references therein). It can also give relevant insight into behaviour of strongly bounded fermion pairs on the optical lattices.

The phase separations involving superconducting states have been evidenced in a~broad range of currently intensely investigated materials including iron-pnictides, cuprates and organic conductors (see for example \cite{AAS2010,PIN2009,RPC2011,XTP2008,UAC2009,SFG2002,MWB2006,PNB2001,KPK2004,CSP2008,TCS2008} and references therein).

In our work we will study a~model which is a~simple generalization of the standard model of a local pair superconductor with on-site pairing (i.~e. the model of hard core bosons on a~lattice \cite{MRR1990,MR1992,MRK1995,BBM2002}) to the case of finite pair binding energy.
The Hamiltonian considered has the following form:
\begin{equation}\label{row:ham1}
\hat{H}=U\sum_{i}{\hat{n}_{i\uparrow}\hat{n}_{i\downarrow}}- I\sum_{\langle i,j\rangle}{\left(\hat{\rho}_i^+\hat{\rho}_j^- + \hat{\rho}_j^+\hat{\rho}_i^-\right)} - \mu\sum_i\hat{n}_i,
\end{equation}
where
\mbox{$\hat{n}_{i}=\sum_{\sigma}{\hat{n}_{i\sigma}}$}, \mbox{$\hat{n}_{i\sigma}=\hat{c}^{+}_{i\sigma}\hat{c}_{i\sigma}$}, \mbox{$\hat{\rho}^+_i=(\hat{\rho}^-_i)^\dag=\hat{c}^+_{i\uparrow}\hat{c}^+_{i\downarrow}$};
$\hat{c}_{i\sigma}$ ($\hat{c}^{+}_{i\sigma}$) denotes the annihilation (creation) operator of an electron with spin \mbox{$\sigma=\uparrow,\downarrow$} at the site $i$,
which satisfy canonical anticommutation relations
\begin{equation}\label{row:antycomutation}
\{ \hat{c}_{i\sigma}, \hat{c}^+_{j\sigma'}\} = \delta_{ij}\delta_{\sigma\sigma'}, \quad
\{ \hat{c}_{i\sigma}, \hat{c}_{j\sigma'}\} = \{ \hat{c}^+_{i\sigma}, \hat{c}^+_{j\sigma'}\} = 0,
\end{equation}
where $\delta_{ij}$ is the Kronecker delta.
\mbox{$\sum_{\langle i,j\rangle}$} indicates the sum over nearest-neighbour sites $i$ and $j$ independently.
$z$ will denote the number of nearest neighbours.
$\mu$ is the chemical potential, depending on the concentration
of electrons:
\begin{equation}\label{row:condn1}
n = \frac{1}{N}\sum_{i}{\left\langle \hat{n}_{i} \right\rangle},
\end{equation}
with \mbox{$0\leq n \leq 2$} and $N$ is the total number of lattice sites. There are two interaction parameters of the model: (i)~the intersite pair hopping interaction $I$, determining the electron pair mobility and responsible  for the long-range superconducting order in the system
and (ii)~the on-site density-density interaction $U$, which contributes (together with $I$) to the pair binding energy by reducing (\mbox{$U>0$}) or enhancing (\mbox{$U<0$}) its value. To simplify our analysis we do not include in Hamiltonian (\ref{row:ham1}) the single electron hopping term ($t_{ij}$) as well as other inter-site interaction terms. This assumption corresponds to the situation when the single particle mobility is much  smaller than the pair mobility and can be neglected.
In the presence of the $t_{ij}$ term, model (\ref{row:ham1}) is called the Penson-Kolb-Hubbard (PKH) model \cite{HD1993,RB1999,JKS2001,CR2001,DM2000,Z2005,MM2004,MC1996} and the effects of \mbox{$t_{ij}\neq0$} will be pointed out in section~\ref{sec:conclusions}.

Model (\ref{row:ham1}) can be considered as a~relatively simple, effective
model of a~superconductor with local electron pairing. Moreover, the knowledge of the \mbox{$t_{ij} = 0$} limit can be used as starting point for a perturbation expansion in powers of the hopping $t_{ij}$ and provides a benchmark for various approximate approaches analyzing the corresponding finite bandwidth models.

In terms of pseudospin charge operators \mbox{$\{\hat{\vec{\rho}}_i\}$} Hamiltonian (\ref{row:ham1}) takes the form of a special XY-model with a~single ion anisotropy in a~transverse field (in the $z$-direction)
\begin{eqnarray}
\hat{H} & = & 2U\sum_i(\hat{\rho}^z_i)^2 -  I\sum_{\langle i,j\rangle}{\left(\hat{\rho}_i^+\hat{\rho}_j^- + \hat{\rho}_j^+\hat{\rho}_i^-\right)} + \nonumber\\
\label{row:ham2}
& - & \bar{\mu} \sum_i \left( 2\hat{\rho}^z_i+1\right) - NU/2,
\end{eqnarray}
where \mbox{$\hat{\rho}^z_i=\frac{1}{2}(\hat{n}_i-1)$} and \mbox{$\bar{\mu}=\mu-U/2$} corresponds to the external magnetic field.
The operators \mbox{$\{\hat{\vec{\rho}}_i\}$} satisfy standard commutation relations for momentum operators:
\begin{equation}
[\hat{\rho}_i^z, \hat{\rho}_j^{\pm}] = \pm \hat{\rho}^\pm_i \delta_{ij}, \qquad [\hat{\rho}_i^{+}, \hat{\rho}_j^{-}] = 2\hat{\rho}_i^{z}\delta_{ij},
\end{equation}
what follows directly from~(\ref{row:antycomutation}).
However, one should stress that \mbox{$\{\hat{\rho}^z_i\}$} takes on four values \mbox{$\{-1/2,0,0,+1/2\}$}, as for \mbox{$S=1$} model with doubly degenerated $0$ value, whereas \mbox{$(\hat{\rho}^+_i)^2=(\hat{\rho}^-_i)^2=0$}, as in the \mbox{$S=1/2$} system.
Condition (\ref{row:condn1}) for the electron concentration reads:
\begin{equation}\label{row:condn2}
\frac{1}{2}(n - 1)= \frac{1}{N}\sum_{i}{\left\langle \hat{\rho}^z_{i} \right\rangle}.
\end{equation}
It means that magnetic field $\bar{\mu}$ is determined by a~fixed value of magnetization $(n-1)/2$.

The interactions $U$ and $I$ will be treated as the effective ones and will be assumed to include all the possible contributions and renormalizations like those coming from the strong electron-phonon coupling or from the coupling between electrons and other electronic subsystems in solid or chemical complexes \cite{MRR1990}. In such a general case arbitrary values and signs of $U$ and $I$ are important to consider. It is notable that formally $I$ is one of the off-diagonal terms of the Coulomb interaction \mbox{$I_{ij}=-(1/2)(ii|e^2/r|jj)$}~\cite{H1963}, describing a part of the so-called bond-charge interaction, and it sign of the \textit{Coulomb-driven} charge exchange is typically negative (repulsive, \mbox{$I<0$}). However, the effective attractive of this form (\mbox{$I>0$}) is also possible~\cite{FH1983,RMR1987,BL1988} and in particular it can originate from the coupling of electrons with intersite (intermolecular) vibrations via modulation of the hopping integral~\cite{FH1983}, or from the on-site hybridization term in generalized periodic Anderson model~\cite{RMR1987,BL1988}.

The ferromagnetic XY order of pseudospins $\hat{\vec{\rho}}_i$ (for \mbox{$I>0$}) corresponds to the SS phase ($s$-pairing superconducting), whereas the antiferromagnetic XY order (for \mbox{$I<0$}) -- to the $\eta$S phase ($\eta$-pairing superconducting). However, in the absence of the external field in the XY-direction conjugated with the order parameter (\mbox{$\Delta=\Delta_{SS} = \frac{1}{N}\sum_i{\langle \hat{\rho}^-_i\rangle}$}, for \mbox{$I>0$} and \mbox{$\Delta_{\eta S} = \frac{1}{N}\sum_i{\exp{(i\vec{Q}\cdot\vec{R}_i)}\langle \hat{\rho}^-_i\rangle} $}, for \mbox{$I<0$}, $\vec{Q}$ is a~half of the smallest reciprocal lattice vector) there is a symmetry between the ferromagnetic and antiferromagnetic cases (except for an obvious redefinition of the order parameter) for lattices consisting of two interpenetrating sublattices (such as for example SC or BCC lattices), thus we restrict ourselves to the \mbox{$I>0$} case in the following. Let us stress that the symmetry is broken by the single electron hopping term \mbox{$t_{ij}\neq0$} (cf. section~\ref{sec:conclusions}).

The first analysis  of the phase diagram of model (\ref{row:ham1}) have been performed by Bari~\cite{B1973} and Ho and Barry~\cite{HB1977} using the variational method in order to examine the instability of the Mott insulator to superconductivity for the special case of the half-filled band (\mbox{$n=1$}). The effects of diagonal disorder  on the critical temperature for \mbox{$U=0$} and \mbox{$n=1$} have been also determined~\cite{WA1987}, arriving at a~satisfactory qualitative interpretation of quite a number of different experiments in amorphous superconductors.

In the analysis we have adopted the variational approach (VA) which treats the on-site interaction $U$ exactly and the intersite interactions $I$ within the mean-field approximation (MFA). We have also derived exact ground state results for a~linear chain (\mbox{$d=1$}). Moreover, at \mbox{$T=0$} several results obtained within the random phase approximation (RPA) and  the spin-wave approximation (SWA) (for \mbox{$d=2$} and \mbox{$d=3$} lattices)  and within the systematic low density expansions (LDE) (for \mbox{$d=3$} lattices) are presented.

Within the VA the phase diagrams of model (\ref{row:ham1}) have been investigated in~\cite{RP1993,R1994} as a function of the electron concentration $n$, but the stability conditions of states with phase separation have not been discussed.

In this paper, we investigate the thermodynamic properties of model Hamiltonian (\ref{row:ham1}) for  arbitrary chemical potential and arbitrary electron concentration. We focus on the states with phase separation, which have not been analyzed till now. Arbitrary values of on-site interaction $U$ are considered and  the range of pair hopping interaction $I$ is restricted to the nearest neighbours.  Our investigation of the general case finds that, depending on the values of the interaction parameters and the electron concentration, the system can exhibit the superconducting ordered and nonordered homogeneous phases as well as the phase separation between them. Transitions between various states and phases can be continuous and discontinuous, what implies existence of tricritical points on the phase diagrams. We present detailed results concerning the evolution of phase diagrams and thermodynamic properties  as a function of interaction parameters, chemical potential $\mu$ and electron concentration $n$.
By taking into account the phase separated states
and presenting several new results for lattices of various dimensions derived beyond VA (rigorous solution for a~chain (\mbox{$d=1$}), RPA for SQ and SC lattices and LDE in \mbox{$d=3$})
our paper substantially 
extends and generalizes the results and conclusions of previous papers concerning the model considered~\cite{RP1993,R1994}.

The paper is organized as follows. In section~\ref{sec:method} we describe the method used in this work. There are also derived explicit formulas for the free energies of homogeneous phases and states with phase separation as well as equations determining the superconducting order parameter  and the chemical potential in homogeneous phases obtained in VA. In section~\ref{sec:GS} we analyze the properties of the system at zero temperature and present exact ground state results for a~chain (section~\ref{sec:GSexact}),  diagrams obtained in the VA (section~\ref{sec:GSMFA}) as well as results obtained in some approximations going beyond VA for \mbox{$d=2,3$} lattices (section~\ref{sec:RPA} and section~\ref{sec:LDE3D}). Section~\ref{sec:kT} is devoted to the study of the finite temperature phase diagrams derived within the VA as a function of $\mu$ (section~\ref{sec:kTmi}) and as a function of $n$ (section~\ref{sec:kTn}). Some particular temperature dependencies of the thermodynamic parameters are discussed in section~\ref{sec:properties}. Finally, section~\ref{sec:conclusions} is devoted to conclusions and supplementary discussion. The appendices report some relevant details:  explicit expressions of site-dependent self-consistent VA equations (\ref{app:sitedependentequations}), particle-hole transformations (\ref{app:sym}), and selected analytical VA results (\ref{app:anal}).


\section{The variational method}\label{sec:method}

The free energy of the system as well as the self-consistent equations for the average number of electrons in the system and the average value of the order parameter are derived within site-dependent VA in Appendix \ref{app:sitedependentequations}.
Restricting our analysis to the nearest neighbours and assuming no spatial variations of the order parameter
the grand potential per site obtained in VA
is given by:
\begin{equation}\label{row:grandpotential}
\omega(\bar{\mu}) = \frac{\Omega}{N} =  -\bar{\mu} + 2I_0|\Delta|^2 - \frac{1}{\beta}\ln (2 Z),
\end{equation}
where
\begin{eqnarray*}
Z = \cosh\left(\beta \sqrt{\bar{\mu}^2+ 4|I_0\Delta|^2} \right) + \exp\left( \beta U /2 \right), \\
\bar{\mu} = \mu - U/2, \quad I_0 = zI, \quad \Delta^*  = \frac{1}{N}\sum_i{\langle \hat{\rho}^+_i\rangle},
\end{eqnarray*}
and $\beta=1/(k_BT)$.
The free energy per site: \mbox{$f=\omega+\mu n$} is derived as
\begin{equation}\label{row:freeenergy}
f(n) =\bar{\mu}(n-1) + (U/2)n + 2I_0|\Delta|^2 - \frac{1}{\beta}\ln (2 Z).
\end{equation}
The grand potential $\omega$ is minimized in the system considered for fixed $\mu$, whereas  the free energy $f$ takes the minimum when $n$ is fixed.

The condition for electron concentration (\ref{row:condn1}) and a~minimization of $\omega$ (or $f$) with respect to the superconducting order parameter $|\Delta|$ lead to the following self-consistent equations (for homogeneous phases):
\begin{eqnarray}
\label{row:MFA1}
\frac{\bar{\mu}\sinh\left( \beta \sqrt{\bar{\mu}^2 + 4 |I_0\Delta|^2} \right)}{Z\sqrt{\bar{\mu}^2 + 4 |I_0\Delta|^2}} &=& n-1, \quad \\
\label{row:MFA2}
|\Delta|\left[ \frac{1}{I_0} - \frac{\sinh\left( \beta \sqrt{\bar{\mu}^2 + 4 |I_0\Delta|^2} \right)}{Z\sqrt{\bar{\mu}^2 + 4 |I_0 \Delta|^2}} \right] & = & 0.
\end{eqnarray}
In the homogeneous phases the double occupancy per site defined as
\mbox{$D = \frac{1}{N}\sum_{i}\left\langle \hat{n}_{i\uparrow}\hat{n}_{i\downarrow}\right\rangle$},  has the following form:
\begin{equation}\label{row:Doccup}
D = \frac{n}{2}\left[ 1 - \frac{1}{n}\frac{\exp{(\beta U/2)}}{Z} \right].
\end{equation}
We also introduce the concentration of locally paired electrons \mbox{$n_p=2D$} and the ratio \mbox{$n_p/n=2D/n$}. Notice that $D$ is different from the condensate density (a~fraction of pairs in the condensate) $n_0$. These quantities differ even in the VA at \mbox{$T=0$}: \mbox{$D=n/2$} and \mbox{$n_0=|\langle \hat{\rho}^+ \rangle|^2=(1/4)n(2-n)$} in the superconducting phase.

Equations (\ref{row:MFA1})--(\ref{row:MFA2})   are solved numerically for \mbox{$T\geq0$} and we obtain $|\Delta|$
and $n$ when $\mu$ is fixed or $|\Delta|$ and $\mu$ when $n$ is fixed.
The superconducting phase (SS) is characterized by non-zero value of $|\Delta|$, whereas \mbox{$|\Delta| =0$} in the non-ordered (normal) phase (NO).

Equation~(\ref{row:MFA2}) is only the necessary condition for an~extremum of (\ref{row:grandpotential}) (or (\ref{row:freeenergy})) thus the solutions of \mbox{(\ref{row:MFA1})--(\ref{row:MFA2})} can correspond to a~minimum or a~maximum (or a~point of inflection) of $\omega$ (or $f$). In addition, the number of minima can be larger than one, so it is very important to find the solution which corresponds to the global minimum of (\ref{row:grandpotential}) (or (\ref{row:freeenergy})).

We say that the solution of (\ref{row:MFA1})--(\ref{row:MFA2}) corresponds to a~metastable phase if it corresponds to a~minimum of the free energy $f$ or grand potential $\omega$ and the stability condition
\begin{equation}\label{row:stabcondition}
\frac{\partial \mu}{\partial n}>0
\end{equation}
is fulfilled.
Otherwise, we say that the phase is unstable. A~stable (homogeneous) phase is a~metastable phase with the lowest free energy (among  all metastable phases and the phase separated states described below).

Phase separation (PS) is a state in which two domains with different electron concentration: $n_+$ and $n_-$ exist in the system
(coexistence of two homogeneous phases). The free energies of the PS states are calculated from the expression:
\begin{equation}\label{row:freeenergyPS}
f_{PS}(n_{+},n_{-}) = m f_{+}(n_{+}) + (1-m) f_{-}(n_{-}),
\end{equation}
where $f_{\pm}(n_{\pm})$ are values of a free energy of two separating phases at $n_{\pm}$ corresponding to the
lowest homogeneous solution for a~given phase, 
$m$ is a fraction of the system with a charge density $n_+$, \mbox{$1-m$} is a~fraction with density $n_-$ (\mbox{$n_+>n_-$}) and \mbox{$mn_+ +(1-m)n_-=n$}. The minimization of (\ref{row:freeenergyPS}) with respect to $n_+$ and $n_-$ yields the equality between the chemical potentials in both domains:
\begin{equation}\label{row:PS1}
\mu_+(n_+)=\mu_-(n_-)
\end{equation}
(chemical equilibrium)
and the following equation (so-called Maxwell's construction):
\begin{equation}\label{row:PS2}
\mu_+(n_+)=\frac{f_{+}(n_{+})-f_{-}(n_{-})}{n_{+}-n_{-}},
\end{equation}
which is equivalent with equality of grand potentials per site in domains: \mbox{$\omega_+(\mu_+)=\omega_-(\mu_-)$}. It implies that the transitions with a~discontinuous change of $n$ in the system considered for fixed $\mu$ can lead to occurrence of the regions of phase separation in the concentration range \mbox{$n_-<n<n_+$} on the diagrams obtained as a function of $n$. In these regions the homogeneous phases can be metastable as well as unstable, depending on the $n$-dependence of $\mu$ (cf. (\ref{row:stabcondition})).
In the PS states the chemical potential \mbox{$\mu=\mu_+(n_+)=\mu_-(n_-)$} is independent of the electron concentration, i.e. \mbox{$\partial \mu/\partial n=0$}.
For more detailed discussion of Maxwell's construction we refer the reader to existing literature, e.~g.~\cite{AS1991,B2004}.

In the model considered only one type of PS states can occur, which is a coexistence of SS and NO phases.

In the paper we have used the following convention. A~second (first) order transition is a~transition between homogeneous phases  with a~(dis-)continuous change of the order parameter at the transition temperature. A~transition between a~homogeneous phase and the PS state is symbolically named as a~``third order'' transition. At this transition a~size of one domain in the PS state decreases continuously to zero at the~transition temperature.
Second order transitions are denoted by solid lines on phase diagrams, whereas dotted and dashed curves denote first order and ``third order'' transitions, respectively.

The phase diagrams obtained are symmetric with respect to half-filling because of the particle-hole symmetry of Hamiltonian (\ref{row:ham1}), so the diagrams will be presented only in the range \mbox{$\bar{\mu} \leq 0$} and \mbox{$0\leq n\leq 1$}. The transformation formulas of various thermodynamic parameters and quantities such as electron concentration, superconducting order parameter, chemical potential and free energy under the electron-hole transformation are collected in Appendix \ref{app:sym}.

The numerical computations have been done in C/C++ with source codes from~\cite{PTVF1992}. The minima have been found by using Brent's method in one dimension and downhill simplex method in multidimensions. We have also implemented Newton-Raphson method using derivative to solve non-linear equations.


\section{The ground state}\label{sec:GS}

Let us consider the ground state (GS) of model (\ref{row:ham1}).

In the absence of $t_{ij}$ term, the parity of the number of particles at each site is a~good quantum number. Thus, one can decompose the eigenspace of $\hat{H}$ into sectors specified by the parity of the occupation number at each site ($n_i$). Within any sector, the system is a collection of \emph{even} segments, within which pairs are confined (the subspace with excluded singly occupied sites) interleaved by \emph{odd} segments (the subspace with excluded doubly occupied sites).

Since \mbox{$I\neq0$}, the interfaces between the \emph{odd}  and \emph{even}  regions will cost energy, so in the thermodynamic limit the ground state will be either a~single infinitely large \emph{even} segment or a~single \emph{odd} segment for fixed $\mu$. Only in the case when the grand potentials per site (it is the chemical potential) of the segments are equal there is possibility of coexistence of both types of segments if \mbox{$N\rightarrow+\infty$} -- it appears as a~first order line on the GS phase diagram.

The situation is different when $n$ is fixed. In such case the PS states can occur in definite ranges of $n$.
The reason for this can be explained as the following. Let the energy of a single boundary (interface) between \emph{odd}  and \emph{even} segments be \mbox{$E_{I}$}. If the number of intersite boundaries will increase
as $N^\gamma$ (where \mbox{$\gamma<1$}) the contribution from them to the total free energy \mbox{$N^\gamma E_I/N =  E_I/N^{1-\gamma}$} disappears in the thermodynamic limit \mbox{$N\rightarrow + \infty$}.

Within the subspace excluding the double occupancy of sites (by electrons for \mbox{$n<1$} or holes for \mbox{$n>1$}) the product \mbox{$\hat{\rho}^+_i\hat{\rho}^-_j=0$} for any \mbox{$i\neq j$}, i.~e. the $I$ term does not contribute. Effective Hamiltonian (\ref{row:ham1}) for this case takes the form \mbox{$\hat{H}_{I}=U\sum_i{\hat{n}_{i\uparrow} \hat{n}_{i\downarrow}} - \mu\sum_i{ \hat{n}_i}$} and
the calculations of the GS grand potential \mbox{$\omega_{NO}=\langle \hat{H}_I\rangle/N$} and the GS energy \mbox{$E_{NO}=\langle \hat{H}_{I}+\mu\sum_i{\hat{n}_i}\rangle/N$} are quite obvious. As a result in the limit \mbox{$\beta\rightarrow+\infty$}  from  (\ref{row:freeenergy}) and (\ref{row:MFA1}) one obtains for fixed $n$ in the NO phase:
\begin{equation}\label{row:GSEXe1}
\begin{array}{lcl}
E_{NO}(n) = 0, &\quad&\textrm{for}\ n<1 \quad (\bar{\mu} = -U/2),\\
E_{NO}(n) = U(n-1), &\quad& \textrm{for}\ n>1 \quad (\bar{\mu} = U/2),\\
E_{NO} = 0 , &\quad& \textrm{for}\ n=1 \quad (\bar{\mu} = 0).
\end{array}
\end{equation}
This GS is infinitely degenerate for \mbox{$N\rightarrow\infty$}. For \mbox{$n\leq1$} (\mbox{$n>1$}), $nN$ \mbox{$((2-n)N)$} sites are occupied by electrons (holes) and the spins of all particles are independent of each other. Inclusion of the intersite spin-spin and density-density interactions can remove this degeneracy and generate various magnetic and charge orderings. The magnetic moments are not suppressed in this state. For \mbox{$n\neq1$} there is no order pattern in the distribution of the particles, whereas for \mbox{$n=1$} Mott phase with one-electron-per-site configuration (with random direction of electron spin per site) occurs.

From (\ref{row:grandpotential}) and (\ref{row:MFA1}) for fixed $\bar{\mu}$ and \mbox{$\beta\rightarrow +\infty$} one obtains:
\begin{equation} \label{row:GSEXomega1}
\begin{array}{lcl}
\omega^a_{NO}(\bar{\mu}) = 0  &\quad& (n=0, \textrm{NO -- empty}), \\
\omega^b_{NO}(\bar{\mu}) = -\bar{\mu}-U/2 &\quad& (n=1, \textrm{NO -- Mott}),\\
\omega^c_{NO}(\bar{\mu}) = -2\bar{\mu} &\quad& (n=2, \textrm{NO -- full}).
\end{array}
\end{equation}
The average occupancy of sites can only be integer for fixed $\bar{\mu}$, what can be intuitively understand if one realizes that \mbox{$\langle \hat{n}_i \rangle=0,1,2$} only and every site is equivalent (there is no interactions between different sites in $\hat{H}_I$).
The state $\omega^a_{NO}$ (NO -- empty) is stable with respect to $\omega^b_{NO}$ (NO -- Mott) if \mbox{$U<-2\bar{\mu}$} (and \mbox{$\bar{\mu}<0$}), whereas \mbox{$\omega^c_{NO}<\omega^b_{NO}$} if \mbox{$U<2\bar{\mu}$} (and \mbox{$\bar{\mu}>0$}).

Within the subspace excluding the single occupancy of sites the eigenvalues of charge operators are \mbox{$\hat{\rho}^z_i=\pm1/2$}, \mbox{$(\hat{\rho}^z_i)^2=1/4$}
for each site. Thus, from (\ref{row:ham2}), the effective Hamiltonian for this case takes the form:
\begin{equation}\label{row:ham3XY}
\hat{H}_{II}= - I\sum_{\langle i,j\rangle}{\left(\hat{\rho}_i^+\hat{\rho}_j^- + \hat{\rho}_j^+\hat{\rho}_i^-\right)} - \bar{\mu}\sum_i{\left(2\hat{\rho}^z_i+1\right)}
\end{equation}
with auxiliary condition (\ref{row:condn2}). $\hat{H}_{II}$ has the form of the XY quantum spin model (\mbox{$S=1/2$})
with an effective external field \mbox{$\bar{\mu}=\mu-U/2$} in the $z$-direction, such that the average magnetization has a fixed value equal to \mbox{$(n-1)/2$} (cf. (\ref{row:condn2})). The GS energy of Hamiltonian (\ref{row:ham3XY}) is derived as
\begin{equation}\label{row:GSEXe2}
E_{SS}(n)=\frac{1}{N}\langle \hat{H}_{II} + \mu \sum_i{\hat{n}_i}\rangle =  E_{XY}(n)+\frac{1}{2}Un,
\end{equation}
where \mbox{$E_{XY}(n) = -\frac{I}{N}\langle\sum_{\langle i,j\rangle}{\left(\hat{\rho}_i^+\hat{\rho}_j^- +
\hat{\rho}_j^+\hat{\rho}_i^-\right)}\rangle$}. The GS grand potential in this case has~the form
\begin{eqnarray}\label{row:GSEXomega2}
\omega_{SS}(\bar{\mu}) = E_{XY}(n(\bar{\mu}))-\bar{\mu}n(\bar{\mu}),
\end{eqnarray}
where it is underlined that $\bar{\mu}$ is an independent variable.

As it follows from previous analyzes~\cite{MR1992,MRK1995} the GS of model (\ref{row:ham3XY}) exhibits the long-range XY order for \mbox{$d\geq2$} dimensional   lattice in the whole concentration range \mbox{$0<n<2$}. For \mbox{$d=1$}, only the short range XY correlations are developed \cite{LSM1961}, nevertheless even arbitrary small interchain interactions can induce the real long-range order also in this case.

The SS phase will be realized if \mbox{$E_{NO}>E_{SS}$} for fixed $n$  (if \mbox{$\omega_{NO}>\omega_{SS}$} for fixed $\bar{\mu}$).
For explicit calculation of the phase boundary between the NO (Eqs.~(\ref{row:GSEXe1}) and (\ref{row:GSEXomega1})) and the SS (Eqs.~(\ref{row:GSEXe2}) and (\ref{row:GSEXomega2})) phases and determination of the GS properties of the SS phase for various dimensions and the electron concentration or the chemical potential one has to find the solution of model (\ref{row:ham3XY}) at \mbox{$T=0$}.

Let us stress that if $n$ is fixed one has to consider also the PS state in which one domain is in the NO phase and the other is in the SS phase and which energy is calculated from (\ref{row:freeenergyPS}). However, the GS energy of the PS (\mbox{SS--NO}) state  can be obtained in the most cases only numerically by minimization of the free energy given by (\ref{row:freeenergyPS}). In the NO domain \mbox{$n_{NO}=1$} (Mott phase with one-electron-per-site configuration), whereas concentration in the SS domain is dependent on $U/I_0$ (and approach used).

    \begin{figure*}
        \centering
        \includegraphics[width=0.32\textwidth]{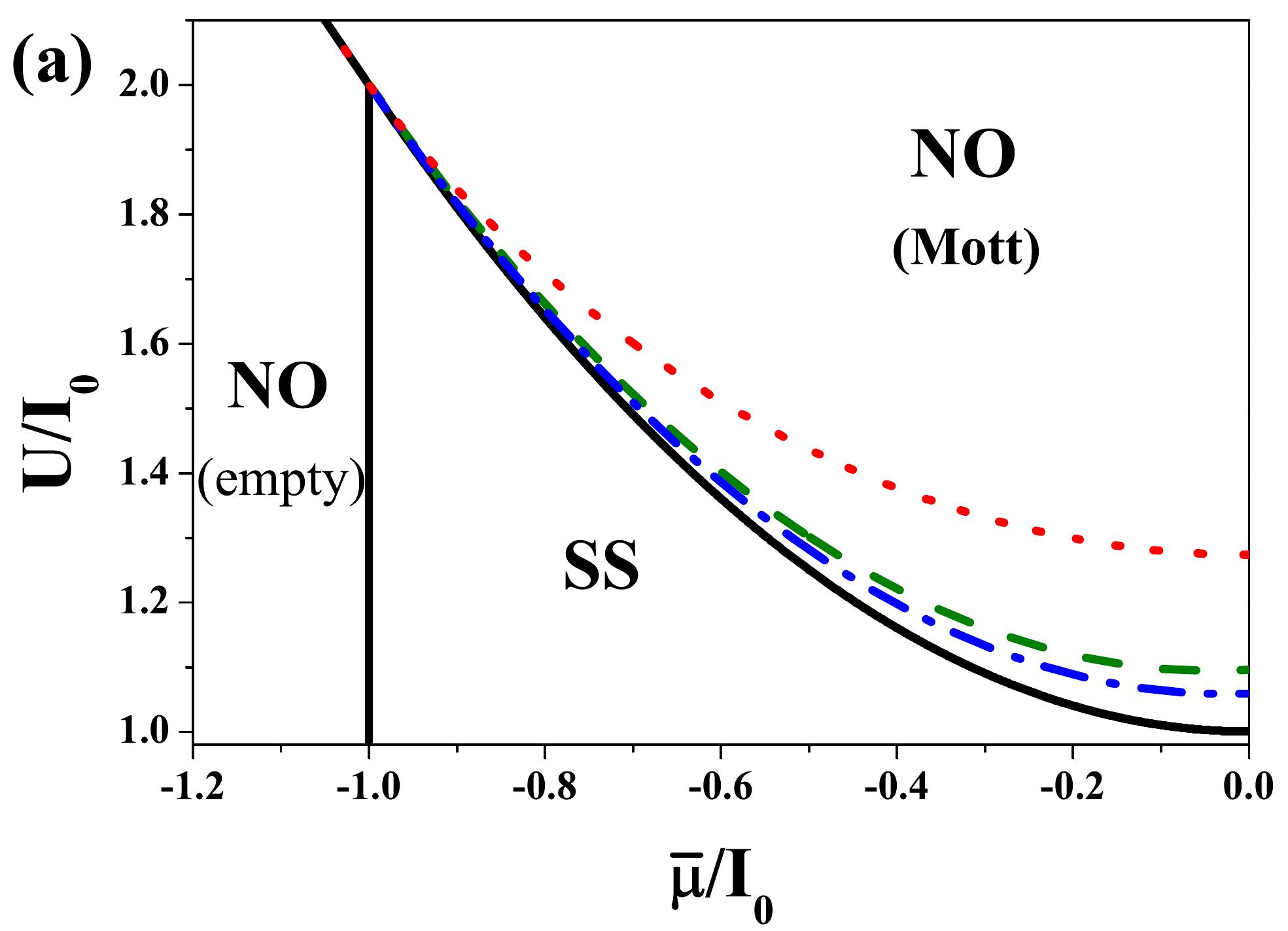}
        \includegraphics[width=0.32\textwidth]{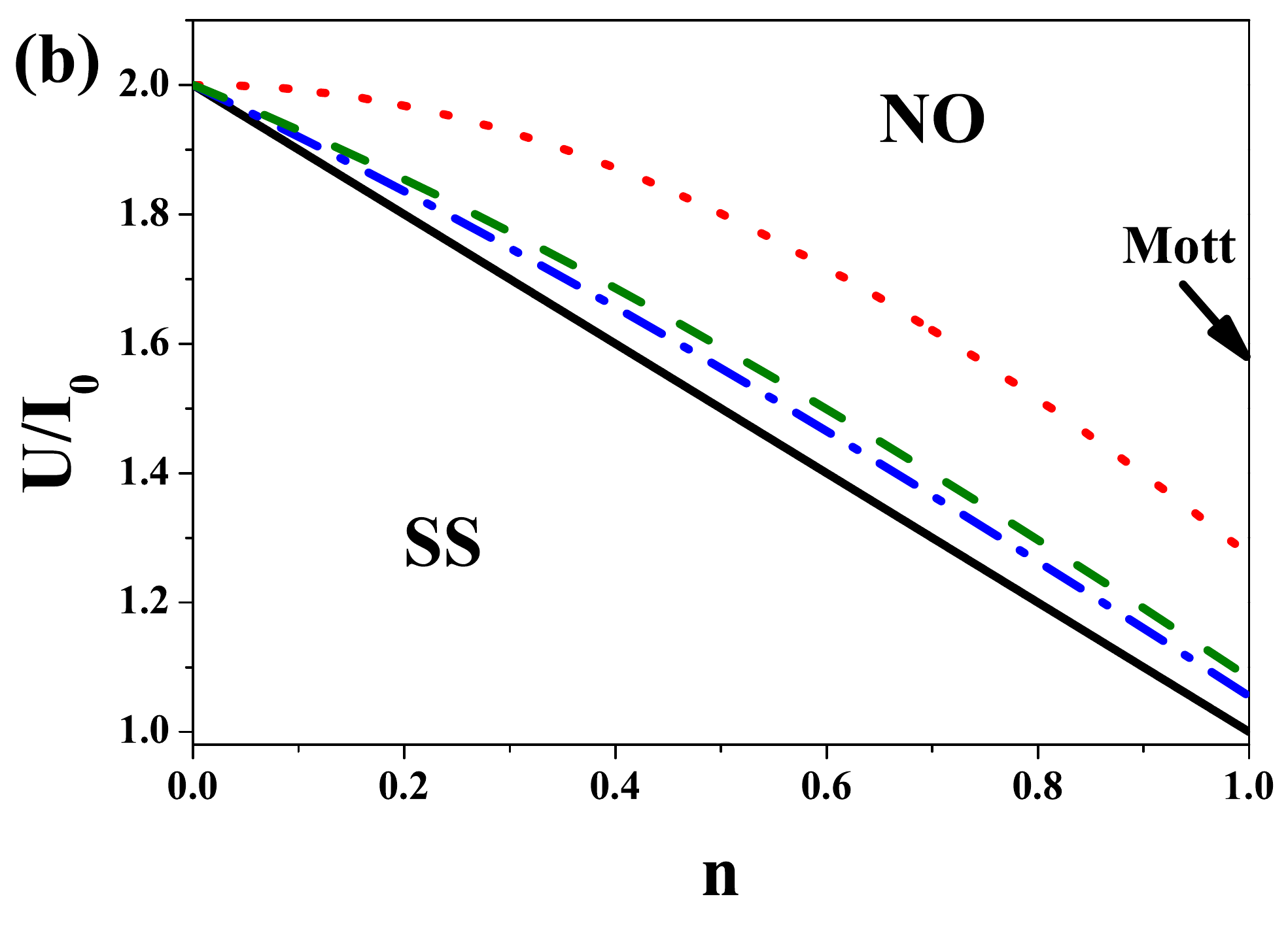}
        \includegraphics[width=0.32\textwidth]{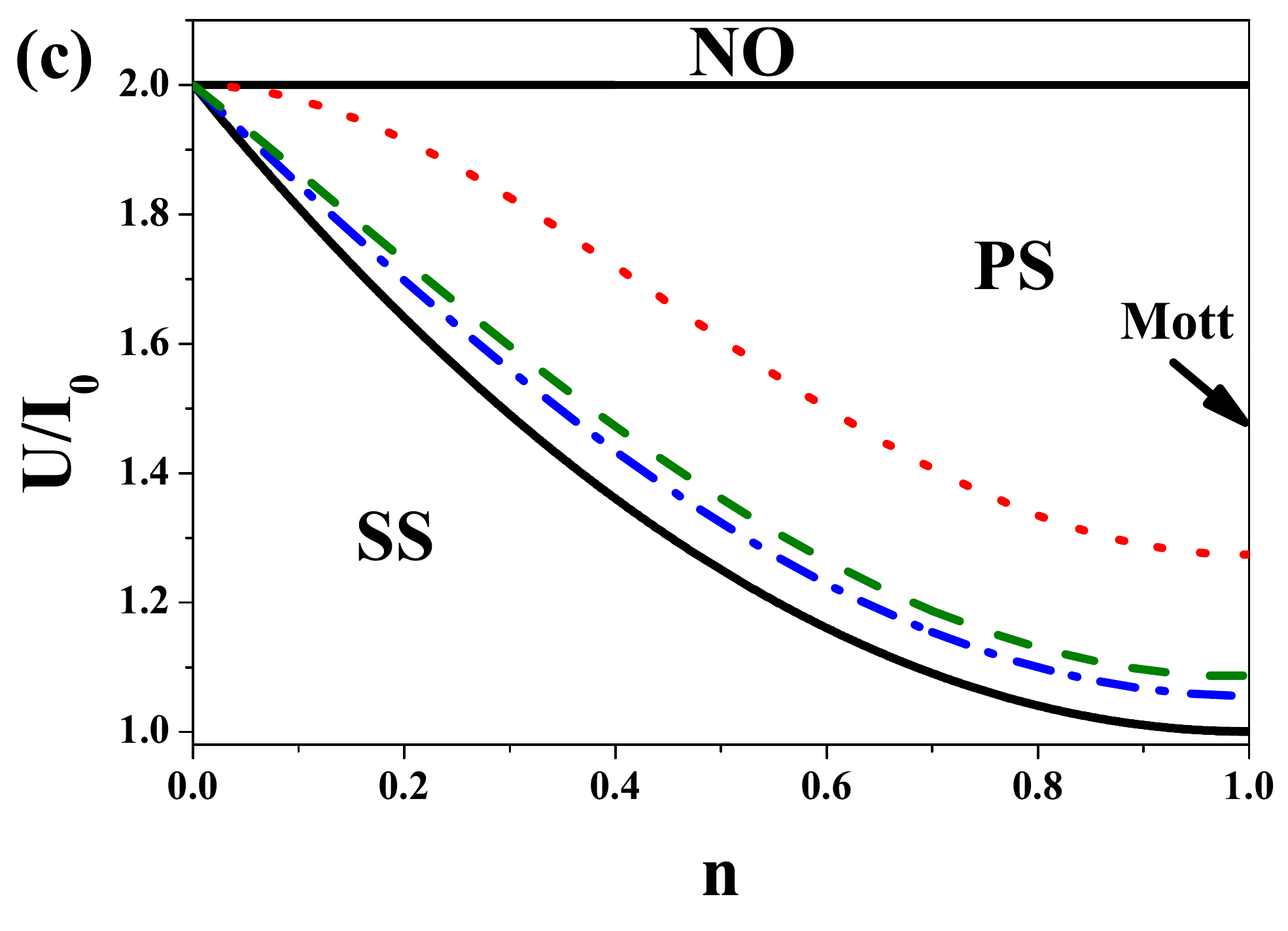}\\
        \caption{(Color online)
        Ground state phase diagrams: (a) as a function of $\bar{\mu}/I_0$, (b) and (c) as a~function of $n$  without and with consideration of the PS states, respectively. Dotted, dashed, dashed-dotted and solid  lines denote the boundaries for \mbox{$d=1$} (exact results), \mbox{$d=2$} (RPA, SQ lattice), \mbox{$d=3$} (RPA, SC lattice), and \mbox{$d\rightarrow+\infty$} (VA result), respectively. At half-filling (\mbox{$n=1$}) the NO (Mott state) is stable above the end of the \mbox{PS--SS} line. The \mbox{NO--SS} transition for \mbox{$\bar{\mu}/I_0=-1$} (panel (a)) is second order, all other transitions between homogeneous phases (panels (a), (b)) are first order.}
        \label{rys:GSdiagrams}
    \end{figure*}


\subsection{Exact results in \mbox{$d=1$}}\label{sec:GSexact}

The exact solutions of model (\ref{row:ham3XY}) for arbitrary $n$ can be derived for \mbox{$d=1$} and \mbox{$d=+\infty$}. For \mbox{$d=1$}, by making use the exact results for the ground state energy of \mbox{$d=1$} XY model in a~transverse field \cite{LSM1961,K1962,N1967,TOK1976} one obtains:
\begin{equation}\label{row:energyXY1d}
E_{XY}^{1D} = - 2|I_0| \frac{1}{\pi} \sin \frac{\pi n}{2},
\end{equation}
where \mbox{$I_0 = 2 I$} (because \mbox{$z=2$}). Notice, that due to a~well known isomorphism between the planar ferromagnet and antiferromagnet for loose packed (alternating) lattices with nearest-neighbour \cite{K1966,SH1973}, the results for $E_{XY}^{1D}$ are the same for both signs of $I$. Thus, from (\ref{row:GSEXe2}) and (\ref{row:energyXY1d}) we have
\begin{equation}
E^{1D}_{SS} = \frac{1}{2}Un - 2|I_0| \frac{1}{\pi} \sin \frac{\pi n}{2}
\end{equation}
and the exact expressions for chemical potential $\mu$ and the superfluid density $j_s$ (the short-range order parameter) are:
\begin{eqnarray}
\label{row:miSSEXGS}
\bar{\mu}^{1D}_{SS} &  =& \frac{\partial E^{1D}_{SS}}{\partial n}  - \frac{1}{2}U =   - |I_0|\cos \frac{\pi n }{2}, \\
j_s^{1D} & = &\left| \langle \hat{\rho}^+_i \hat{\rho}^-_{i+1} \rangle \right| = \frac{1}{\pi}\sin \frac{\pi n }{2},
\end{eqnarray}
where $i+1$ denotes the nearest neighbour of the $i$-site. For \mbox{$I_0>0$} (SS case) ferromagnetic correlations between XY  components of neighbouring pseudospins are present \mbox{$\langle \hat{\rho}^+_i \hat{\rho}^-_{i+1} \rangle>0$}, whereas for \mbox{$I_0<0$} ($\eta$S case) antiferromagnetic correlations occur \mbox{$\langle \hat{\rho}^+_i \hat{\rho}^-_{i+1} \rangle<0$}. From (\ref{row:miSSEXGS}) one obtains that the SS can exist in the range \mbox{$-1<\bar{\mu}/|I_0|<1$}. Notice that stability condition (\ref{row:stabcondition}) is always fulfilled in the SS phase.

The GS phase diagrams obtained are shown in figure~\ref{rys:GSdiagrams}: as a function of $\bar{\mu}$ (panel (a)), and as a function of $n$ without (panel (b)) and with (panel (c)) consideration of the PS state. At half-filling there is no phase separation and the NO - Mott phase is stable for \mbox{$U/I_0>4/\pi$}.
The boundaries between different states corresponding to the exact results in \mbox{$d=1$} are denoted by dotted lines (if they are different from those obtained in VA).
Notice that for homogeneous phases \mbox{$\partial \mu/\partial n >0$} in the ranges of the PS state occurrence.


\subsection{Variational approximation (\mbox{$d\rightarrow+\infty$})}\label{sec:GSMFA}

For \mbox{$1<d<\infty$} there are no exact solutions for model (\ref{row:ham3XY}) available (except of the LDE for \mbox{$d=3$} -- see \cite{MR1992,MRK1995,RP1993}) and the approximate approaches are necessary to find $E_{SS}$. The simplest of these approaches is the VA which at \mbox{$T=0$} gives (the \mbox{$\beta\rightarrow+\infty$} limit of (\ref{row:freeenergy})--(\ref{row:MFA2})
for \mbox{$\Delta\neq0$})
\begin{eqnarray}
E_{SS} = \frac{1}{2}Un- \frac{1}{2}|I_0| n (2-n), \\
|\Delta|^2 =j_s= \frac{1}{4} n (2-n),
\end{eqnarray}
and
\begin{eqnarray}
\label{row:miSSVAGS}\bar{\mu}_{SS} =  - |I_0|(1-n).
\end{eqnarray}
Equation (\ref{row:miSSVAGS}) implies that the SS phase can exist in the range \mbox{$-1<\bar{\mu}/|I_0|<1$}.
Notice that \mbox{$n_p =2D = n$}, which means that all electrons in the system are locally paired in the SS phase at \mbox{$T=0$}, whereas
the finite value of the depletion \mbox{$n'=n-2n_0$} (\mbox{$n_0 = |\Delta|^2$}) at \mbox{$T=0$} is due to the hard-core effect~\cite{MR1992}.

The GS energy of the PS state can be obtained by minimization of (\ref{row:freeenergyPS}). In the NO domain \mbox{$n_{NO}=1$} (Mott phase), whereas concentration in the SS domain is dependent on $U/I_0$ and \mbox{$n_{SS}=1\pm\sqrt{U/I_0-1}$} (\mbox{$1\leq U/I_0 \leq 2$}). It corresponds to the first order \mbox{SS-NO} boundary on the $U/I_0$~vs.~$\bar{\mu}/I_0$ GS diagram determined by equation \mbox{$(\bar{\mu}/I_0)^2+1=U/I_0$}.

The GS phase diagrams obtained in VA are shown in figure~\ref{rys:GSdiagrams}. The VA phase boundaries are denoted by solid lines.
Similarly as in section~\ref{sec:GSexact} one finds that \mbox{$\partial \mu/\partial n >0$} for homogeneous phases in the ranges of the PS state occurrence.

For hypercubic lattice of the dimension \mbox{$d=\infty$} the VA becomes the exact theory for model (\ref{row:ham1}) (as well as for model (\ref{row:ham3XY})).

\subsection{Random phase approximation (\mbox{$d=2,3$})}\label{sec:RPA}

Going beyond VA for \mbox{$1<d<+\infty$} a~reliable approach is the self-consistent RPA~\cite{MR1992,MRK1995}. This approach has been proven to be a~very good approximation scheme in problems  of quantum magnetism and it fully takes into account quantum fluctuations, which can be of crucial importance for the considered system for \mbox{$d\leq3$}. The GS energy of model (\ref{row:ham3XY}) is given by (\ref{row:GSEXe2}), where \mbox{$E^{RPA}_{XY}(n)=-\frac{1}{N}\sum_{\vec{k}} I_{\vec{k}} \langle \hat{\rho}^+_{\vec{k}} \hat{\rho}^-_{\vec{k}}\rangle$}. $\hat{\rho}^\pm_{\vec{k}}$ and $I_{\vec{k}}$ denote the space-Fourier transform of the charge operators and the pair hopping integral, respectively, and the chemical potential is determined by \mbox{$(n-1)/2=(1/N)\sum_i\langle\hat{\rho}^z_i\rangle$}

The energy $E^{RPA}_{XY}$ is obtained by means of Green's functions using the spectral theorem, and by applying the approximation for the longitudinal correlation functions \cite{MRK1995}. At \mbox{$T=0$} this yields
\begin{eqnarray}
E^{RPA}_{XY} = & - & \sin^2 \theta \left[\frac{R^2I_0}{2}+ \frac{R}{2N}\sum_{\vec{k}} I_{\vec{k}} \frac{B_{\vec{k}}}{E_{\vec{k}}} \right]  \\
& -& (1+\cos^2\theta)\frac{R}{2N}\sum_{\vec{k}} I_{\vec{k}} \frac{A_{\vec{k}}}{E_{\vec{k}}}, \nonumber
\end{eqnarray}
where \mbox{$A_{\vec{k}}=R\epsilon^0_{\vec{k}}+B_{\vec{k}}$}, \mbox{$B_{\vec{k}} = RI_{\vec{k}}\sin^2\theta$}, \mbox{$E_{\vec{k}}=\sqrt{A_{\vec{k}}^2-B_{\vec{k}}^2}$} is the collective excitation spectrum, \mbox{$\epsilon^0_{\vec{k}}=2(I_0-I_{\vec{k}})$}, $\theta$ is given by \mbox{$\cos^2\theta = (n-1)^2/R^2$} and the length of pseudospin $R$ is given as a~solution of the self-consistent equation
\begin{eqnarray}
R^{-1} = \frac{1}{N}\sum_{\vec{k}}\frac{A_{\vec{k}}}{E_{\vec{k}}} = 1+2\psi_0, \\
\label{row:RPApsi}
2\psi_0 = \frac{1}{N} \sum_{\vec{k}} \left[ \frac{\epsilon^0_{\vec{k}}+I_{\vec{k}}sin^2\theta}{ \sqrt{(\epsilon^0_{\vec{k}})^2+ 2I_{\vec{k}} \epsilon^0_{\vec{k}}\sin^2\theta}} - 1\right].
\end{eqnarray}
The result corresponding to the SWA (being the lowest order expansion of above equation) is \mbox{$R=1-2\psi_0$} with $2\psi_0$ determined by (\ref{row:RPApsi}).

For the SS phase the spectrum $E_{\vec{k}}$ is gapless and linear in $|\vec{k}|$ for small $\vec{k}$. One gets \mbox{$E_{\vec{k}\rightarrow0}= |\vec{k}|\mathfrak{s}$} with  the sound velocity $\mathfrak{s}$ given by \mbox{$\mathfrak{s}=2 R I a\sqrt{z}\sin\theta$} (\mbox{$\hbar = 1$}).

The chemical potential in the SS phase in the RPA is given by \mbox{$\bar{\mu}_{SS}^{RPA}=I_0(n-1)$}, which is the same as in VA.
The quantum corrections to this VA result for $\bar{\mu}_{SS}$ can be evaluated as in \cite{MRK1995} and one obtains:
\begin{equation}
\bar{\mu}_{SS}=2I_0 \left[\frac{n-1}{2}-\frac{\cos \theta}{2N} \sum_{\vec{k}} \frac{\gamma_{\vec{k}}(1-\gamma_{\vec{k}})}{\epsilon_{\vec{k}}}\right],
\end{equation}
where \mbox{$\epsilon_{\vec{k}}=\sqrt{(1-\gamma_{\vec{k}})^2+\gamma_{\vec{k}}(1-\gamma_{\vec{k}})\sin^2\theta}$}, \mbox{$\gamma_{\vec{k}}=I_{\vec{k}}/I_0$}.

One should notice that RPA and SWA give very similar result for the model (\ref{row:ham3XY}) in the GS~\cite{MRK1995,BBM2002}.
The resulting phase boundaries for \mbox{$d=2$} square (SQ) lattice and \mbox{$d=3$} simple cubic (SC) lattice  are shown in figure~\ref{rys:GSdiagrams} and denoted by dashed and dashed-dotted lines, respectively.

\subsection{Low-density expansion for \mbox{$d=3$} lattices}\label{sec:LDE3D}

In \cite{MR1992,MRK1995} several rigorous results concerning the GS properties of the model of hard-core charged bosons have been derived for various lattices in \mbox{$d=3$} using a~systematic LDE based on the knowledge of the exact two-body scattering amplitude. Transforming the results of section~VI~B from \cite{MRK1995} into our problem one gets for \mbox{$n\rightarrow0$} the following exact expressions. The GS energy of SS phase $E_{SS}^{LDE}$ is given by
\begin{eqnarray}
E_{SS}^{LDE} & = & \frac{1}{2}Un+E_{XY}^{LDE} =  \\
 & = & \frac{1}{2}Un - 2I_0 \left[\frac{1}{2}n - \frac{1}{4}\alpha n^2 - \frac{\gamma z^{(3/2)}}{30\pi^2}(2\alpha n)^{(5/2)}\right], \nonumber
\end{eqnarray}
where \mbox{$\alpha=1/C$} is the exact scattering length for the considered case, \mbox{$C=(1/N)\sum_{\vec{k}}(1-I_{\vec{k}}/I_0)^{-1}$} is the Watson integral for a~given lattice, \mbox{$\gamma=V_0/a^3$}, $V_0$ is the volume of the unit cell, $a$ is the lattice constant. In particular, for SC: \mbox{$C=1.51638$}, \mbox{$z=6$}, \mbox{$\gamma=1$}; for BCC: \mbox{$C=1.3932$}, \mbox{$z=8$}, \mbox{$\gamma=1/2$}; and for FCC: \mbox{$C=1.3446$}, \mbox{$z=12$}, \mbox{$\gamma=1/4$}.

The chemical potential has a form
\begin{equation}
\bar{\mu}^{LDE}_{SS}=-I_0\left[1-\alpha n - \frac{\gamma z^{(3/2)}}{3\pi^2} \alpha^{5/2}(2n)^{3/2}\right]
\end{equation}
and the sound velocity $\mathfrak{s}$ is derived by
\begin{equation}
\hbar^2\mathfrak{s}^2=4I_0Ia^2n(2-n)\alpha \left[ 1+\alpha^{3/2} \frac{2\gamma}{\sqrt{2}\pi^2}z^{3/2}n^{1/2}\right].
\end{equation}

The phase diagrams obtained within LDE are in very good agreement with RPA results (for SC lattice) in the range \mbox{$n<0.2$}.
The homogeneous SS phase boundary derived within LDE is  moved little downwards in comparison with the RPA ones.
An analysis of LDE results obtained for various cubic lattices in \mbox{$d=3$} shows that the region of the SS phase stability shrinks with increasing number of the nearest neighbours in the lattice.


\begin{figure}
    \centering
    \includegraphics[width=0.49\textwidth]{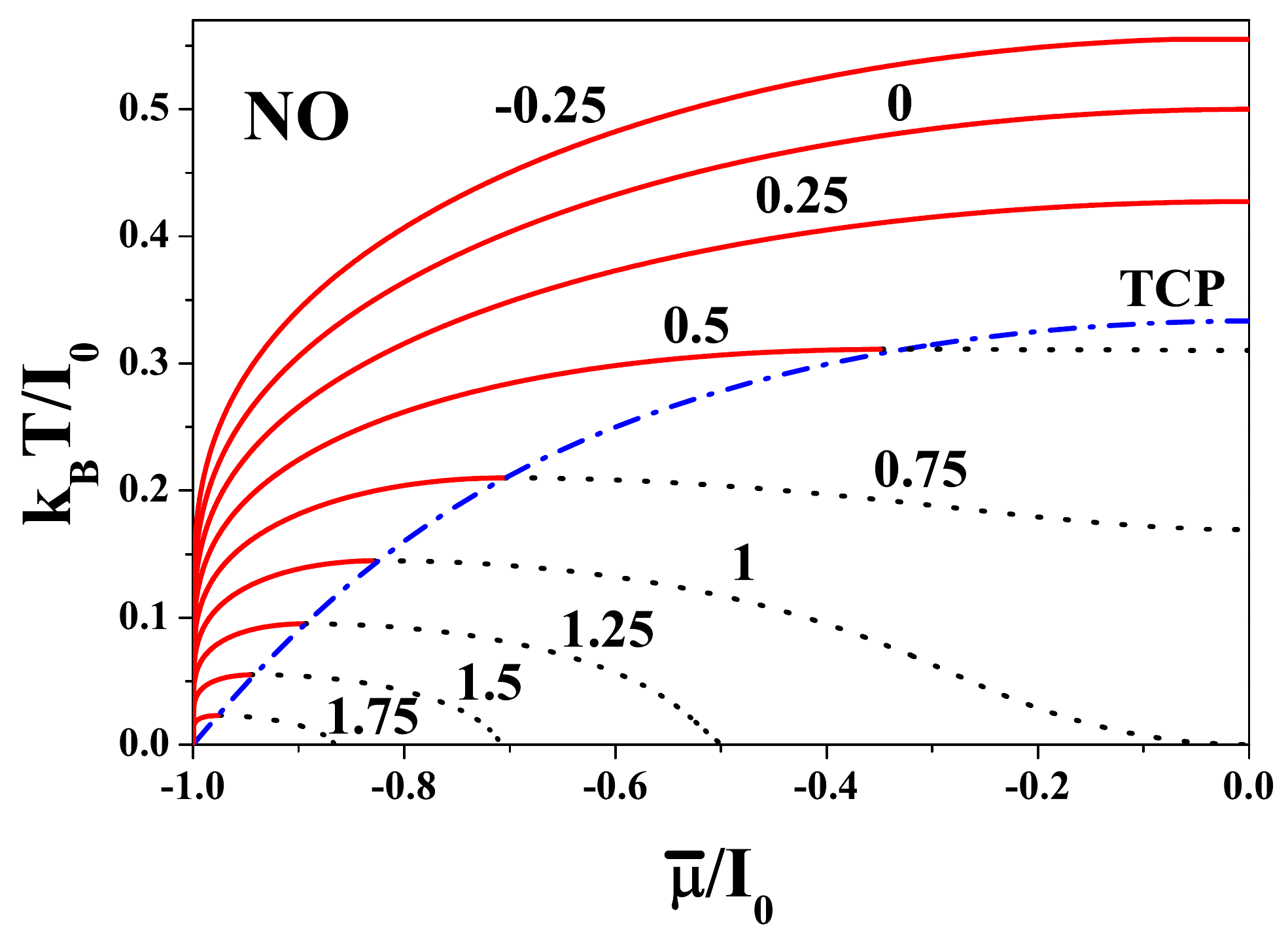}
    \caption{(Color online)
    The SS--NO transition temperature as a~function of $\bar{\mu}/I_0$ for increasing values of $U/I_0$ (numbers above the curves). Solid and dotted lines indicate second order and first order transitions, respectively. The dashed-dotted curve (labeled as TCP) is a~projection of the tricritical point line on the $k_BT/I_0$~vs.~$\bar{\mu}/I_0$ plane.}
    \label{rys:PSkTvsmi}
\end{figure}
\begin{figure}
    \centering
    \includegraphics[width=0.49\textwidth]{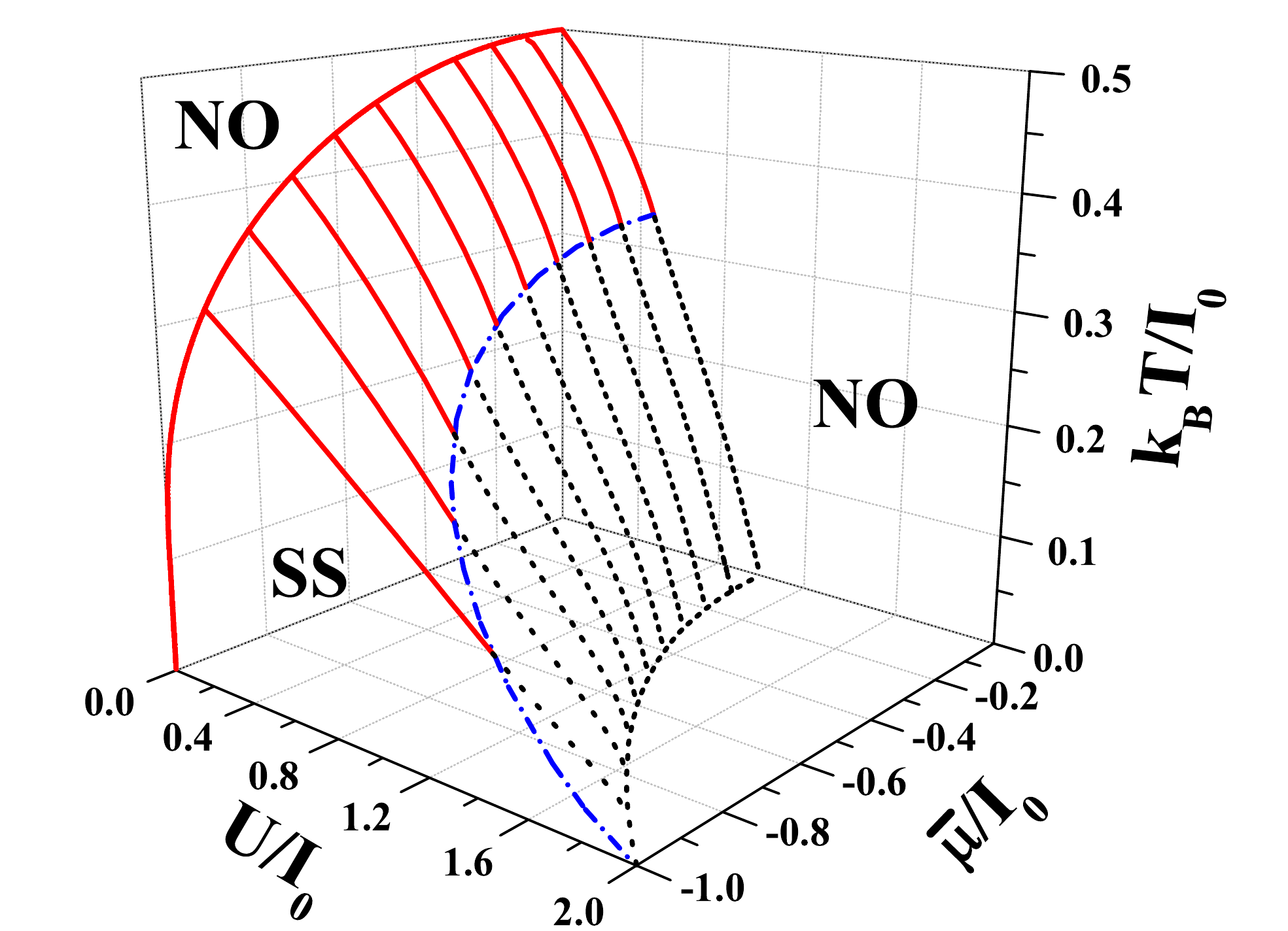}
    \caption{(Color online)
    Finite temperature  phase diagram plotted as a function of $\bar{\mu}/I_0$  and $U/I_0$. Solid and dotted lines indicate second order and first order transitions, respectively. The tricritical point line is denoted by dashed-dotted line.}
    \label{rys:PSkTvsmi3D}
\end{figure}


\subsection{Summary of the GS results}

The general structure of the GS phase diagrams derived is similar for all  lattice dimensionalities (\mbox{$d\geq1$}) and approaches used.
There are two sequences of phase transitions with increasing $U/I_0$ for fixed $n$:
\begin{itemize}
\item[(i)]{SS $\rightarrow$ PS $\rightarrow$ NO (for \mbox{$n\neq 1$}),}
\item[(ii)]{SS $\rightarrow$ NO (Mott) (for \mbox{$n=1$}).}
\end{itemize}

The effects of quantum fluctuations increase with decreasing lattice dimensionality and for a~linear chain their influence on phase diagram is the most prominent. Thus the  region of the homogenous SS phase stability decreases with increasing $d$. In \mbox{$d\rightarrow+\infty$} the quantum fluctuations are suppressed.

One should also notice that the transition PS-NO at \mbox{$U/I_0=2$} is independent of $n$ and dimensionality of the lattice. The PS region expands with increasing dimensionality.


\section{Finite temperature phase diagrams derived with VA}\label{sec:kT}


\subsection{Analysis for fixed $\mu$}\label{sec:kTmi}

\begin{figure*}
    \centering
    \includegraphics[width=0.32\textwidth]{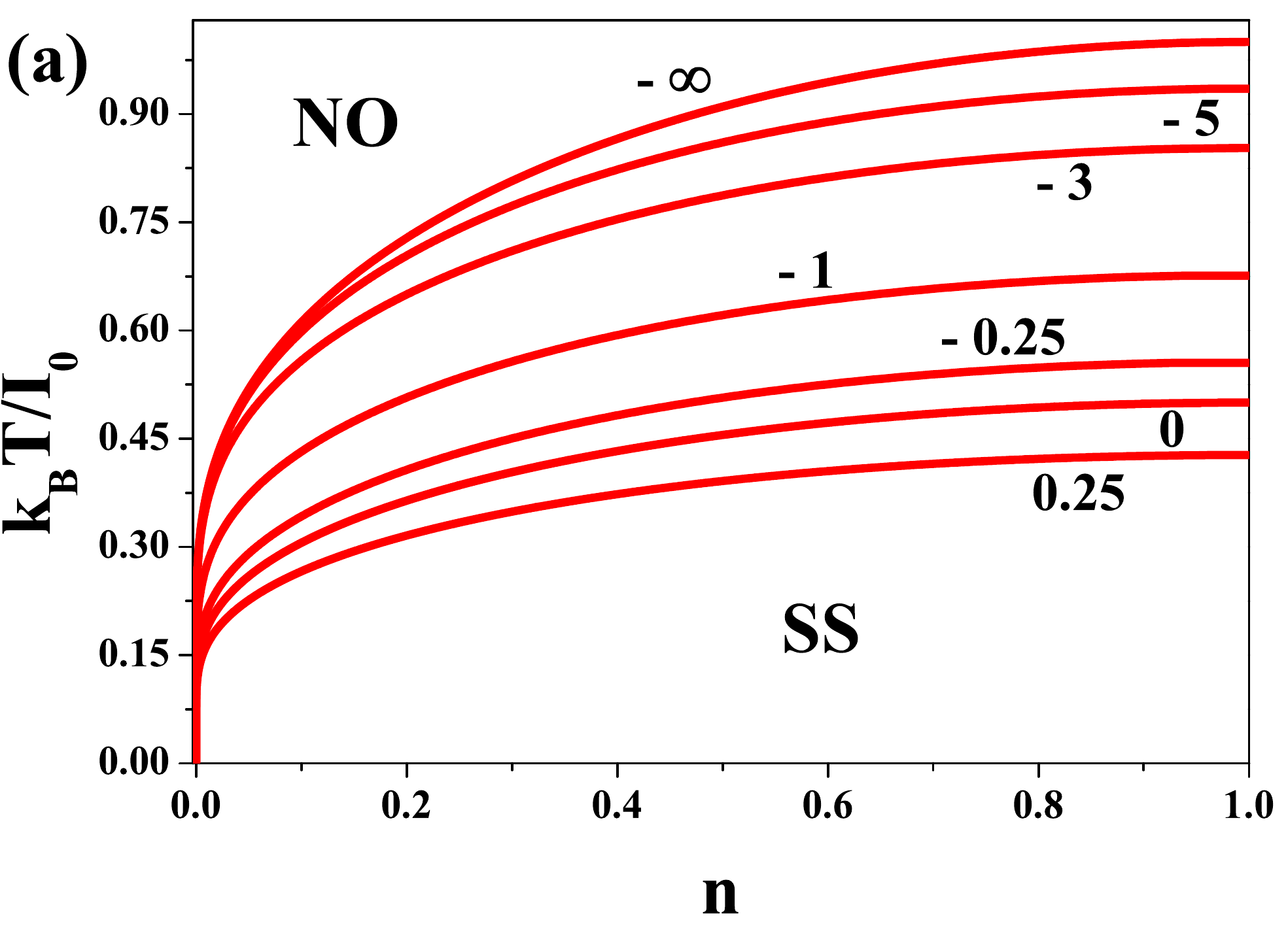}
    \includegraphics[width=0.32\textwidth]{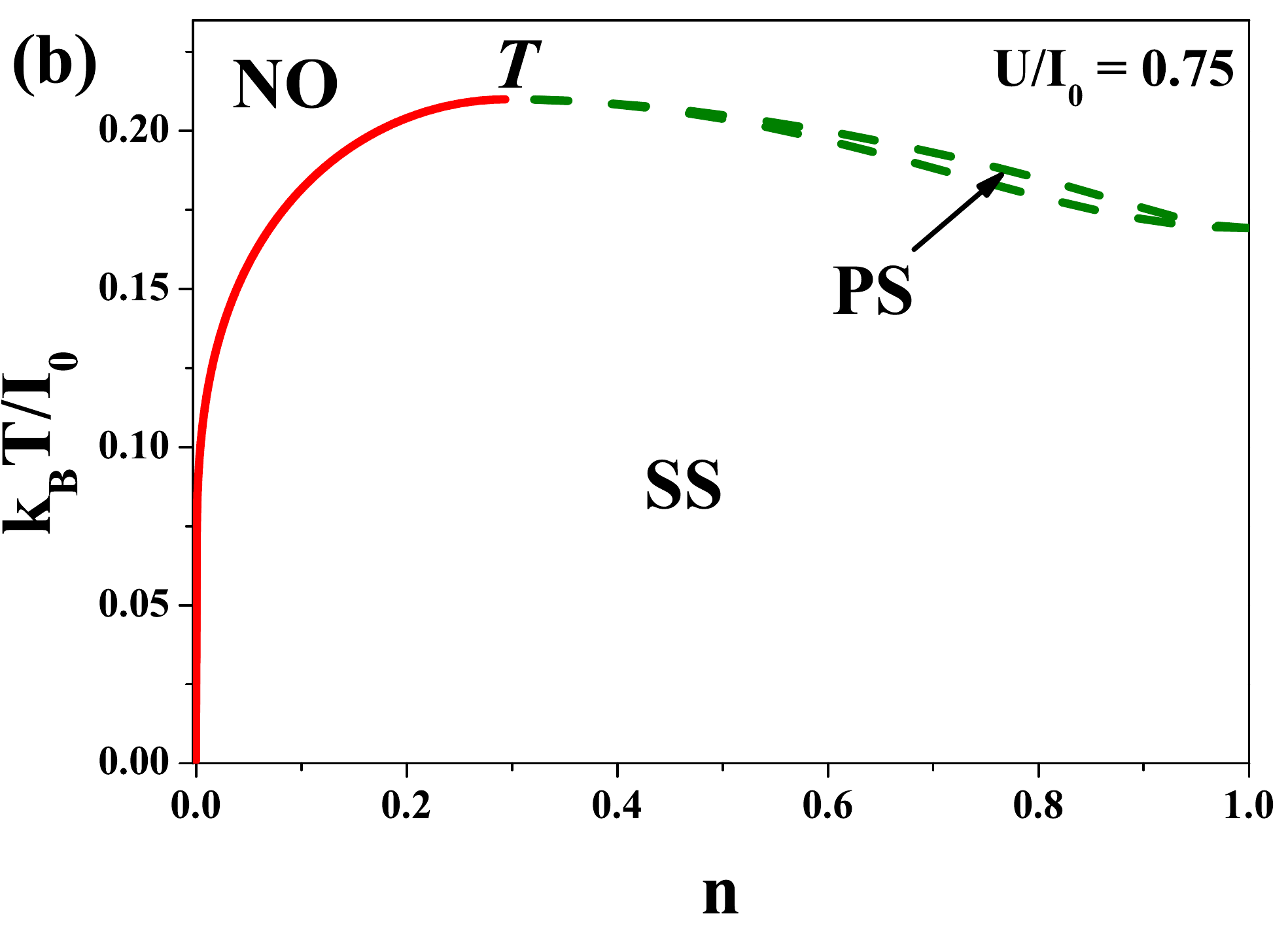}
    \includegraphics[width=0.32\textwidth]{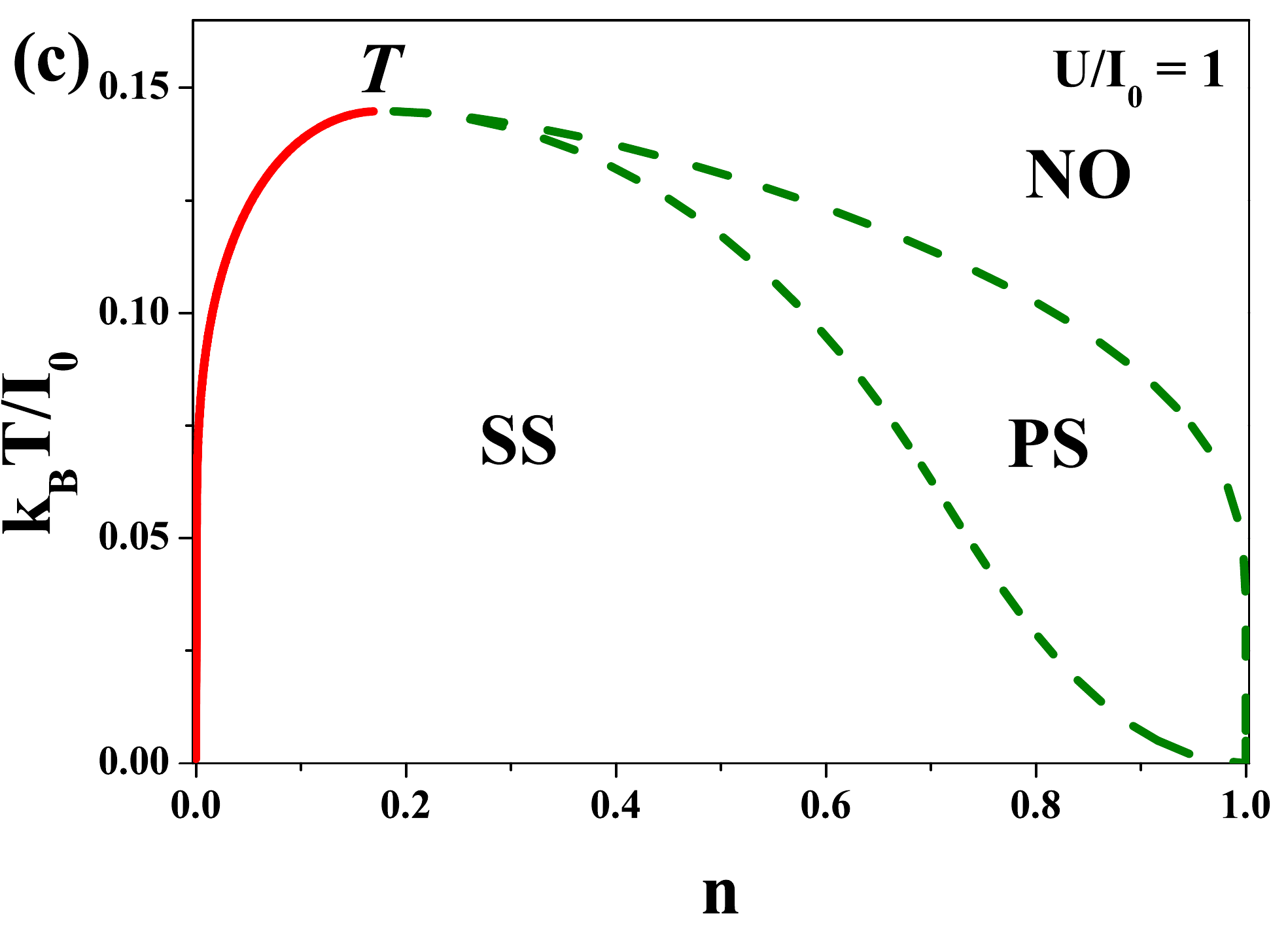}\\
    \includegraphics[width=0.32\textwidth]{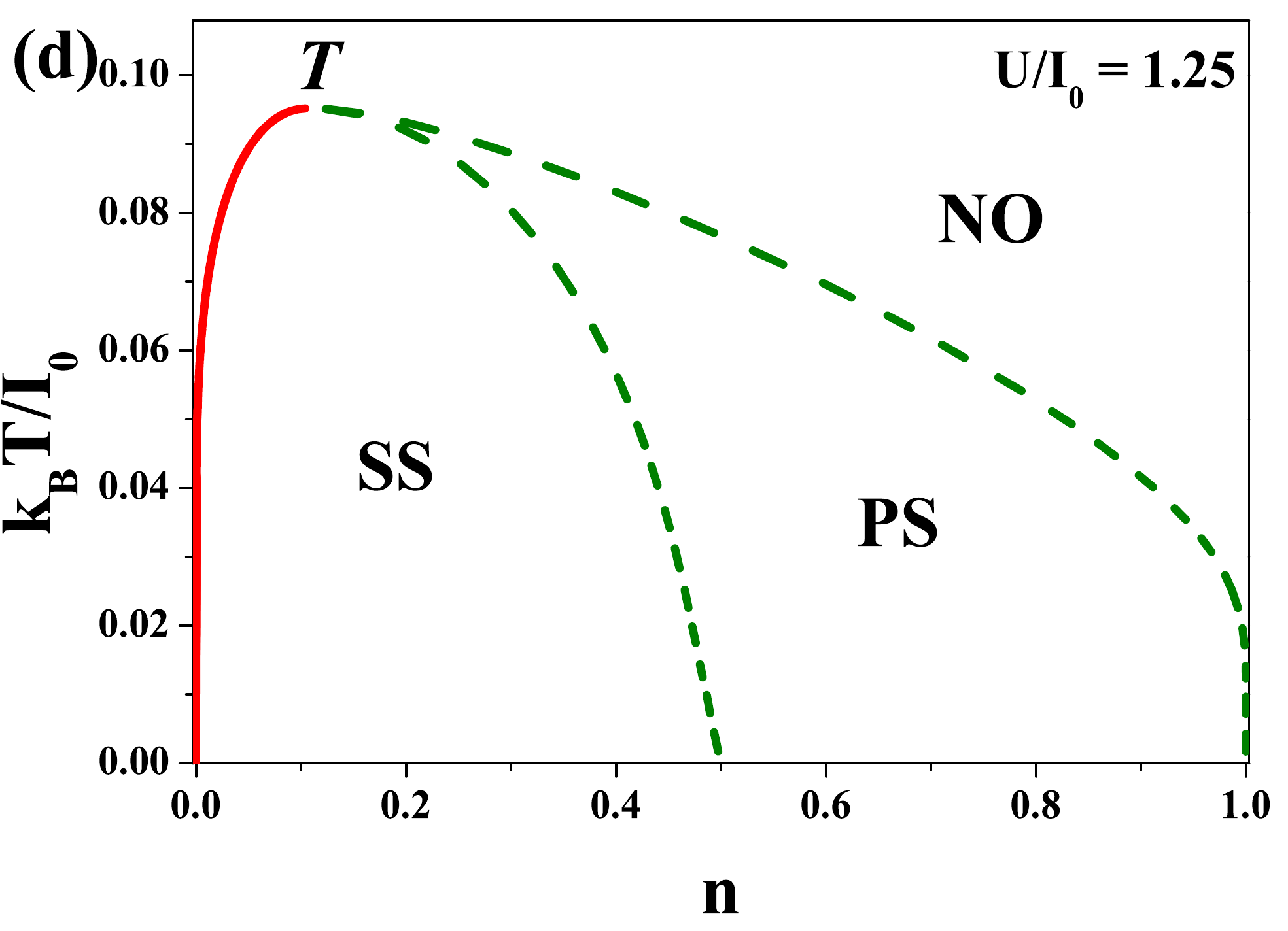}
    \includegraphics[width=0.32\textwidth]{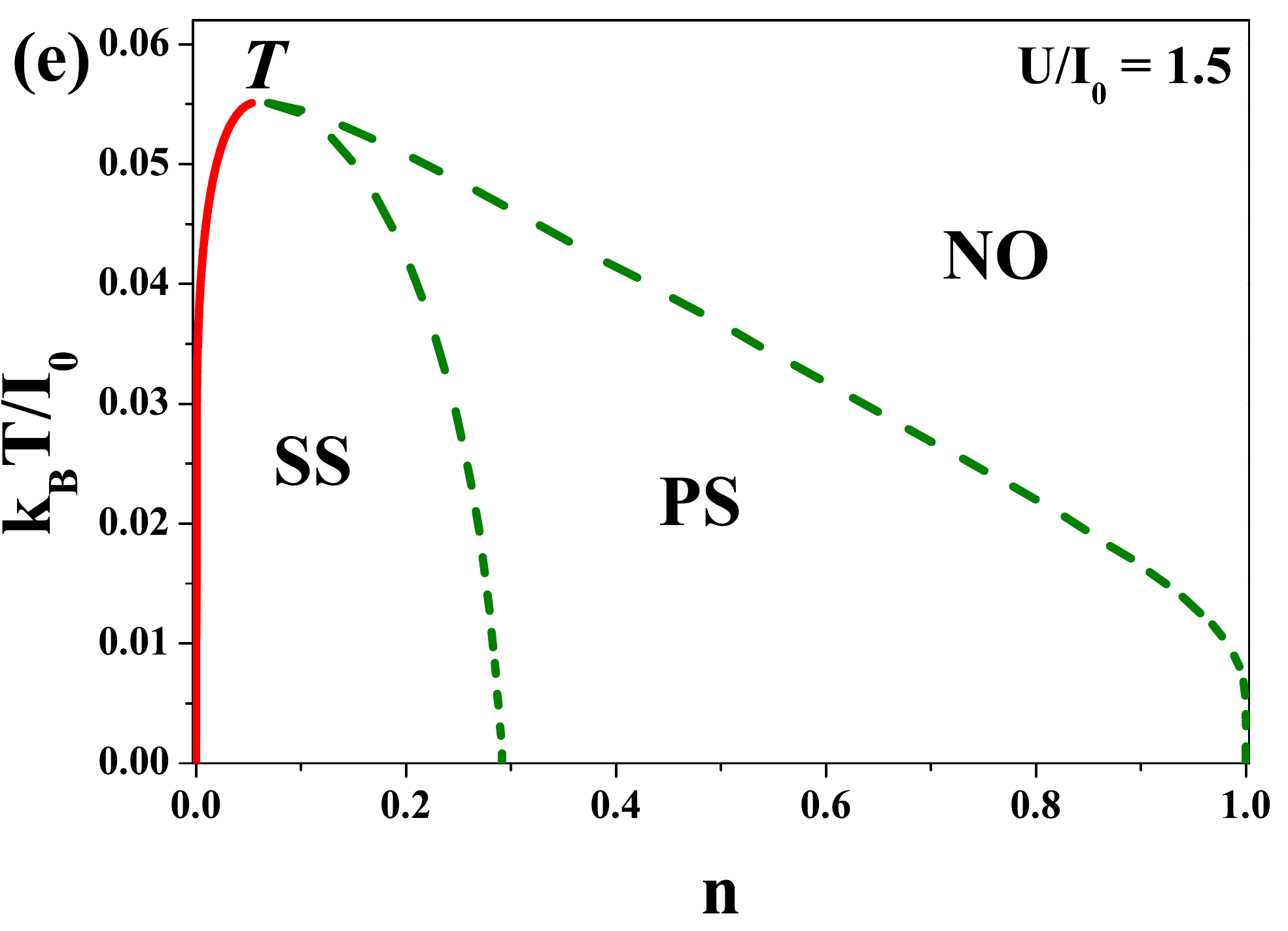}
    \includegraphics[width=0.32\textwidth]{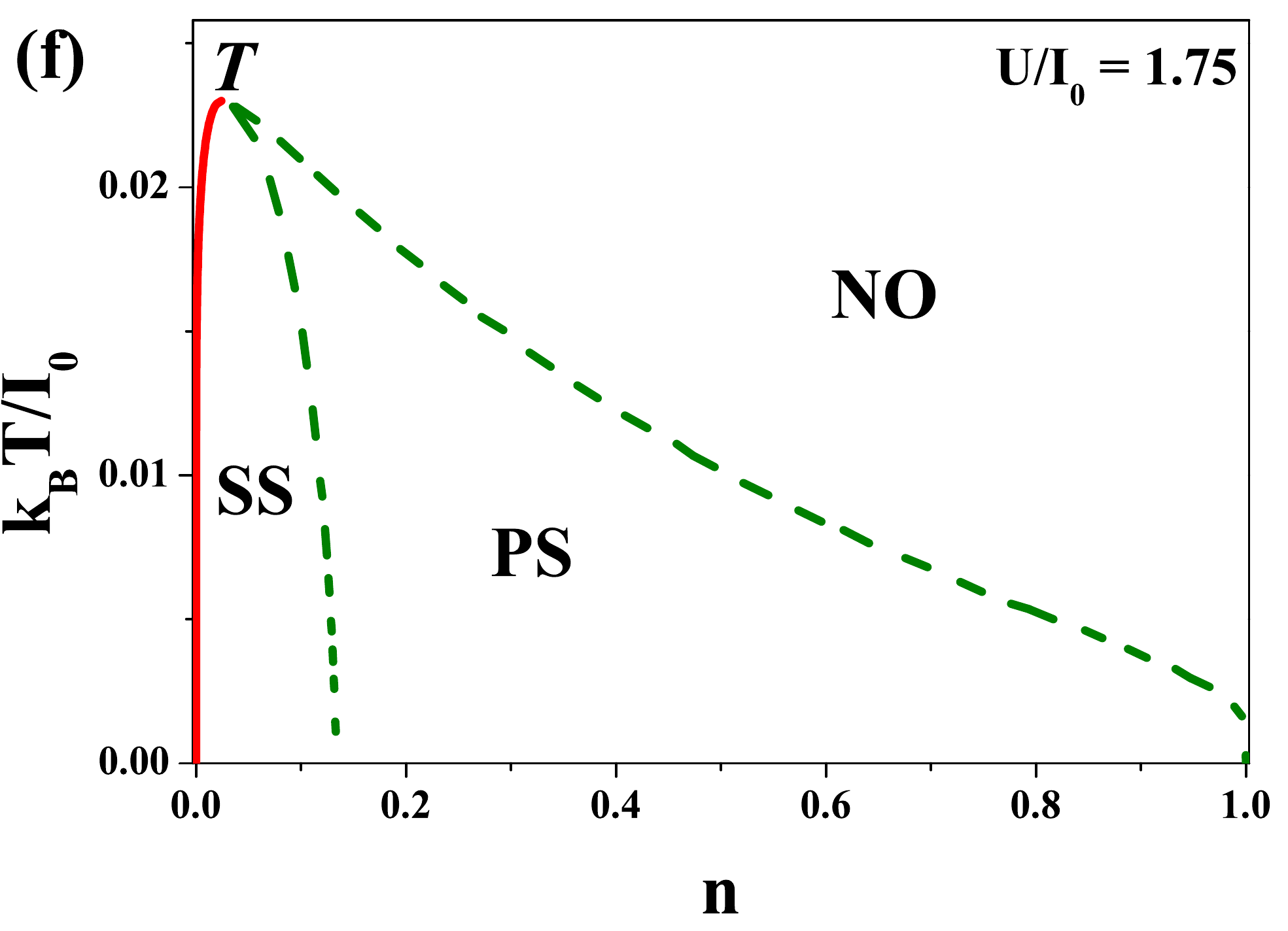}
    \caption{(Color online)
    Phase diagrams $k_BT/I_0$~vs.~$n$ for different values of $U/I_0$ (as labeled). Solid and dashed lines indicate second order and ``third order'' boundaries, respectively. $\mathbf{T}$ denotes the tricritical point.}
    \label{rys:PSkTvsn}
\end{figure*}

All phase transition boundaries, necessary to construct the complete phase diagram for fixed $\mu$, have been obtained numerically by solving the set of equations (\ref{row:MFA1})--(\ref{row:MFA2}) and comparing the grand potential (\ref{row:grandpotential}) for the solutions found. The resulting $k_BT/I_0$~vs.~$\bar{\mu}/I_0$ phase diagrams obtained
for various fixed values of the on-site interaction $U/I_0$ are shown in figure~\ref{rys:PSkTvsmi}. The overall behaviour of the system as a function of $\bar{\mu}$ is demonstrated in figure~\ref{rys:PSkTvsmi3D}.

For \mbox{$U/I_0\leq(2/3)\ln 2$} only the second order  \mbox{SS--NO} transitions occur with increasing temperature. The transition  temperature is maximal for \mbox{$U\rightarrow - \infty$}, \mbox{$\bar{\mu}=0$} and it decreases monotonically with increasing $U/I_0$ and $|\bar{\mu}|/I_0$.

The most interesting is the range  \mbox{$(2/3)\ln 2 < U/I_0 <2$}. In this range the \mbox{$\bar{\mu}$-dependence} of the \mbox{SS--NO} transition temperature becomes non-monotonic and with increasing $U/I_0$ the maximum of the transition temperature moves towards higher values of $|\bar{\mu}|/I_0$.
In this range of the on-site interaction, the order of the \mbox{SS--NO} transition changes from second order to first order with increasing of  $\bar{\mu}/I_0$.  A~\emph{tricritical point} (TCP) is connected with the change of transition order.

For \mbox{$2<U/I_0<+\infty$} only the NO phase is stable at any \mbox{$T\geq0$}.

As it will be showed in section~\ref{sec:kTn} the first order phase boundaries are associated with the existence of the PS states in definite ranges of $n$ if the system is considered for fixed $n$.


\subsection{Analysis as a function of  $n$}\label{sec:kTn}

The transition temperatures have been determined by self-consistent solving of  \mbox{(\ref{row:MFA1})--(\ref{row:MFA2})} and \mbox{(\ref{row:PS1})--(\ref{row:PS2})} and comparing free energies (\ref{row:freeenergy}) and (\ref{row:freeenergyPS}).

The concentration dependencies of (i)~the second order \mbox{SS--NO} transition temperatures and (ii)~the ``third order'' \mbox{SS--PS} transition temperatures can be also easily obtained from the diagrams as a function of $\bar{\mu}$, because of a~simple relation \mbox{$\bar{\mu}_{SS}=I_0(n-1)$} between the electron concentration and the chemical potential in SS phase. The transitions (i) and (ii) for fixed $n$ correspond to the second order and first order \mbox{SS--NO} transitions for fixed $\mu$, respectively. However, there is no simple correspondence for the \mbox{PS--NO} transitions (cf. (\ref{row:miNO})).

Examples of the $k_BT/I_0$~vs.~$n$ phase diagrams evaluated for various values of the on-site interaction $U/I_0$ are shown in figure~\ref{rys:PSkTvsn}.

For \mbox{$U/I_0\leq(2/3)\ln 2$} the PS states do not occur and the obtained phase diagrams are the same as those derived in \cite{RP1993}. The transition between homogeneous SS and NO phases taking place with increasing temperature is second order for arbitrary $n$. The \mbox{SS--NO} transition  temperature is maximal for \mbox{$U\rightarrow - \infty$}, \mbox{$n=1$} and it decreases monotonically with increasing $U/I_0$ and decreasing $n$ (for \mbox{$n\leq 1$}) -- cf. figure~\ref{rys:PSkTvsn}a.

In the range \mbox{$(2/3)\ln 2 < U/I_0 <2$} with increasing $U/I_0$ the maximum of the \mbox{SS--NO} transition temperature moves towards lower concentrations (for \mbox{$n<1$}). In a~definite range of $U/I_0$ and $n$ the PS state \mbox{SS--NO} is stable. For \mbox{$1<U/I_0<2$} the PS state extends from the ground state (cf. figures~\ref{rys:PSkTvsn}d, e, f and figure~\ref{rys:GSdiagrams}c), whereas for \mbox{$(2/3)\ln 2<U/I_0\leq1$} it is stable only at finite temperatures (cf. figures~\ref{rys:PSkTvsn}b, c). The critical point for the phase separation (denoted as $\mathbf{T}$, which is a~\emph{tricritical point}, TCP) lies on the end of the \mbox{SS--NO} second order line.

Between \mbox{$U/I_0=2$} and \mbox{$U/I_0\rightarrow + \infty$} only the NO phase is stable at any
\mbox{$T\geq 0$} and there are no transitions.

\begin{figure*}
    \centering
    \includegraphics[width=0.32\textwidth]{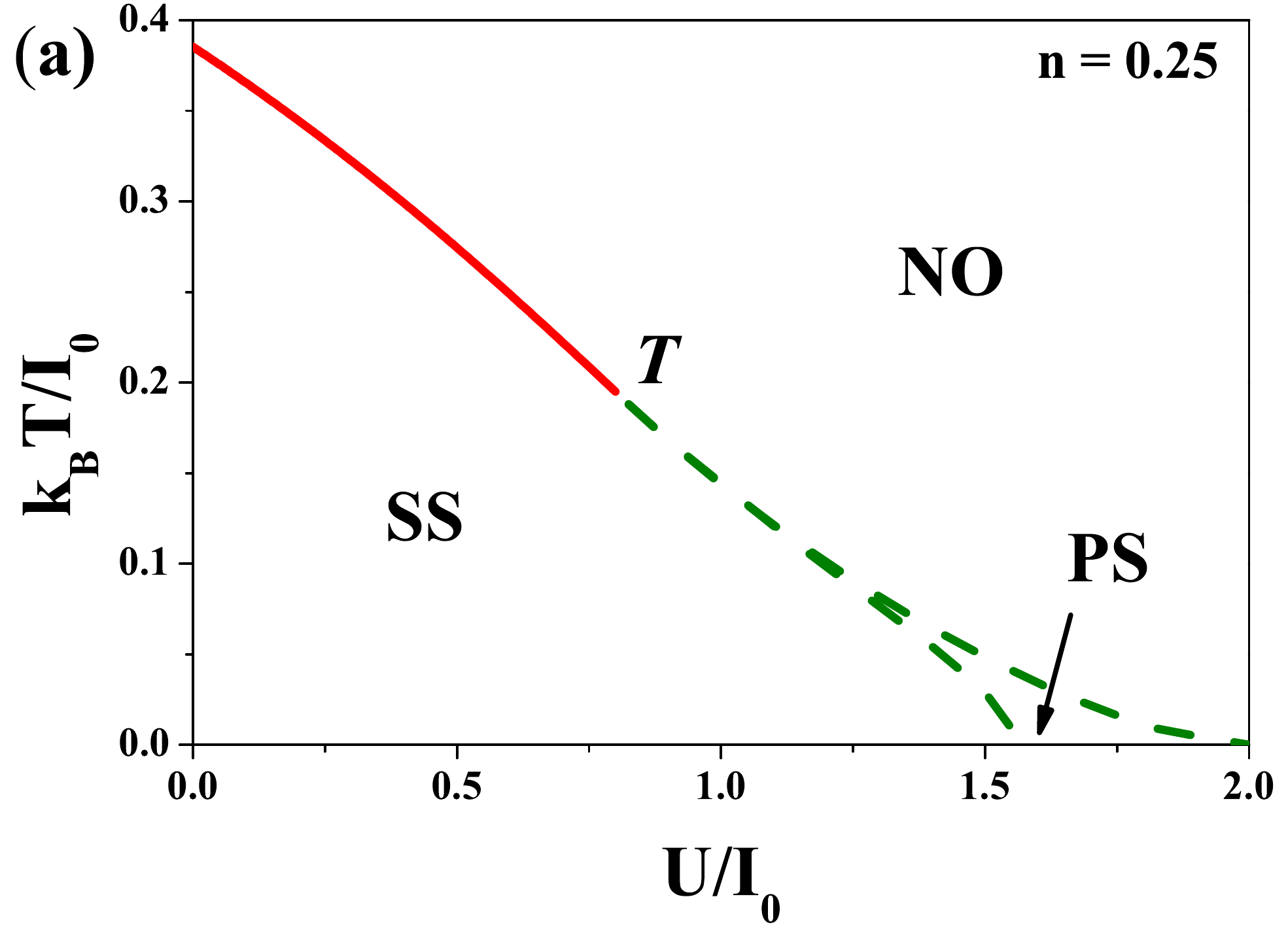}
    \includegraphics[width=0.32\textwidth]{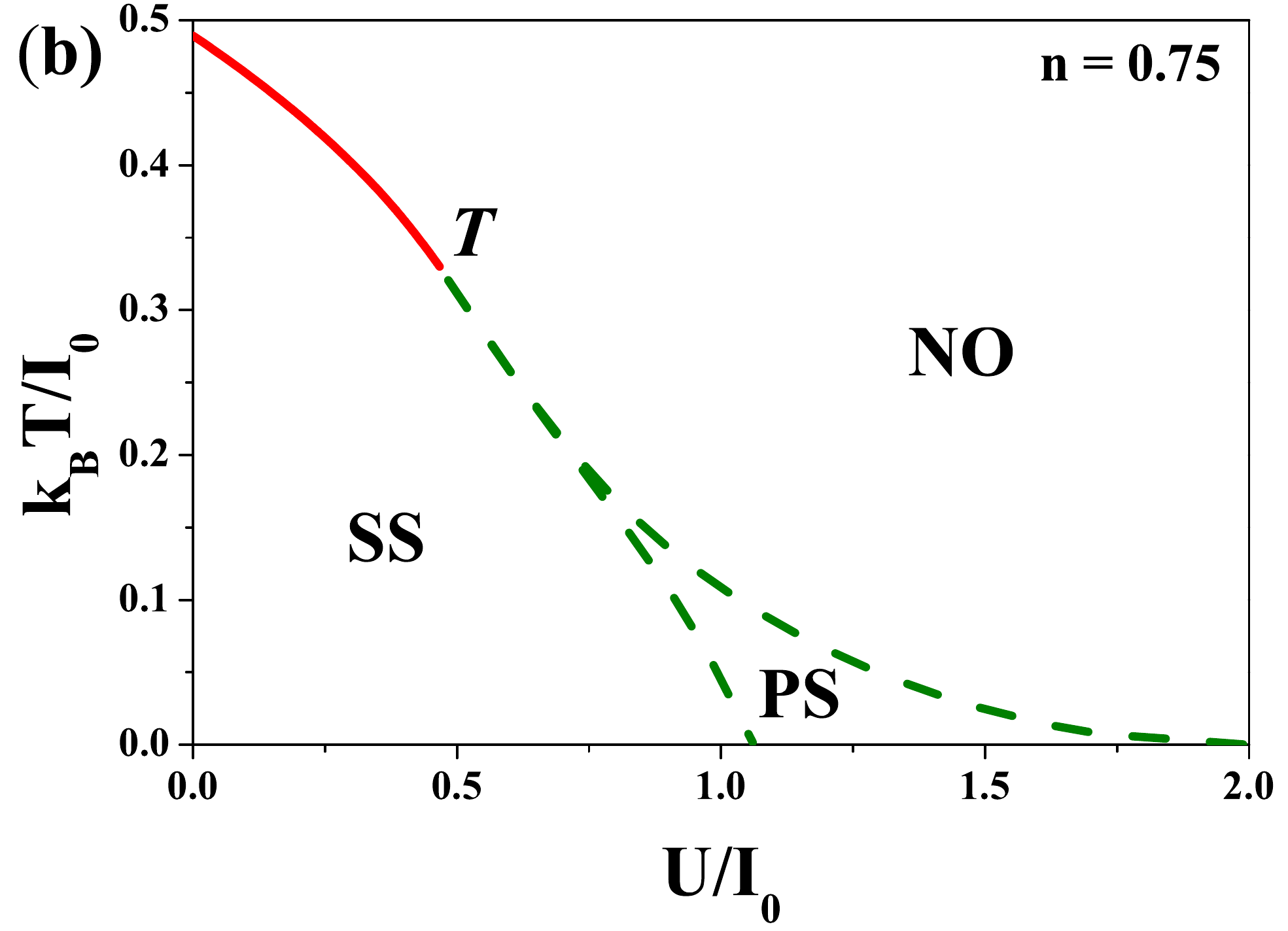}
    \includegraphics[width=0.32\textwidth]{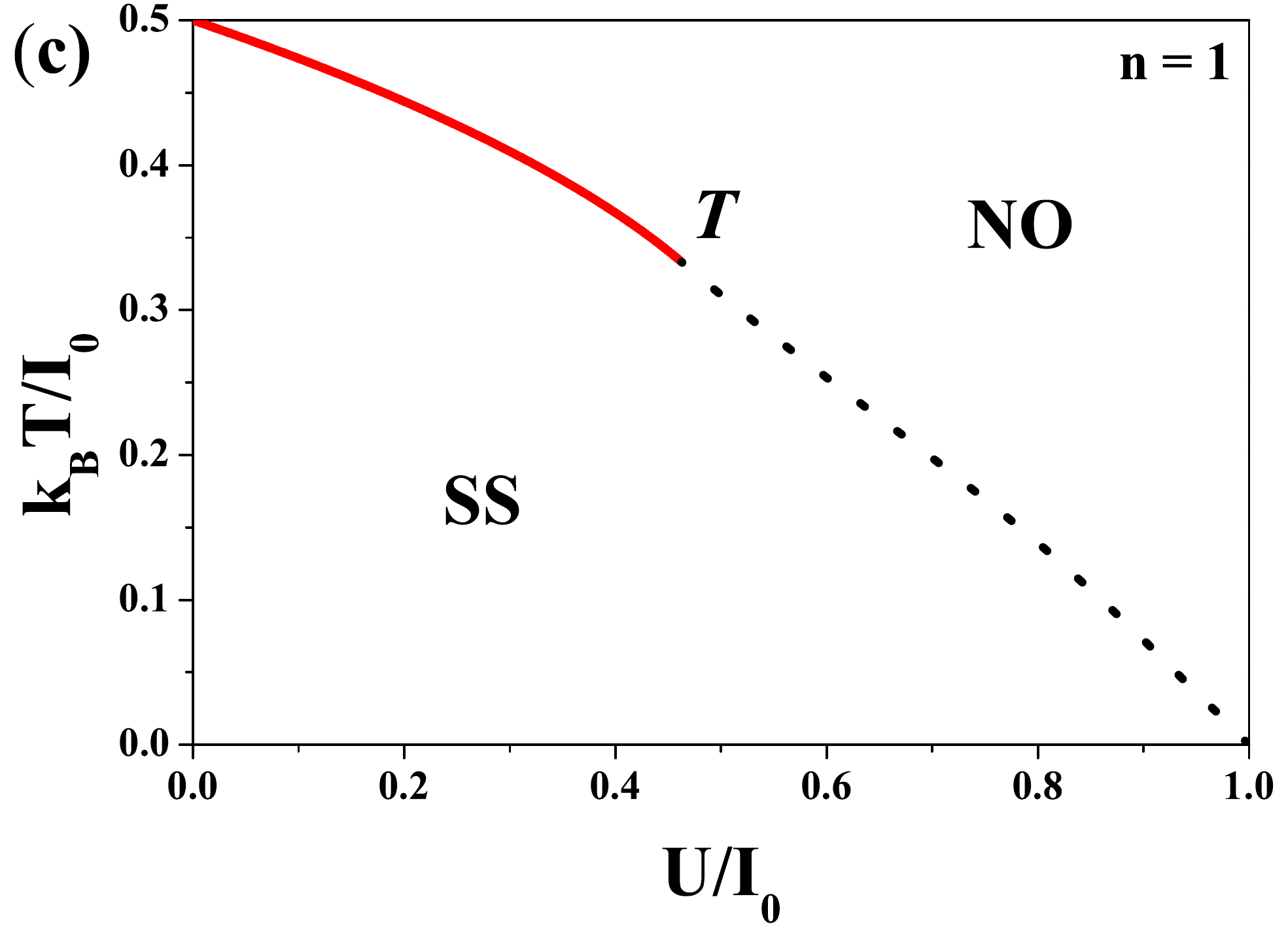}
    \caption{(Color online)
    Phase diagrams $k_BT/I_0$~vs.~$U/I_0$ for \mbox{$n=0.25$} (a), \mbox{$n=0.75$} (b), and \mbox{$n=1$} (c). Solid and dashed lines indicate second order and ``third order'' boundaries, respectively. $\mathbf{T}$ denotes the tricritical point.}
    \label{rys:PSkTvsU}
\end{figure*}
\begin{figure*}
    \centering
    \includegraphics[width=0.32\textwidth]{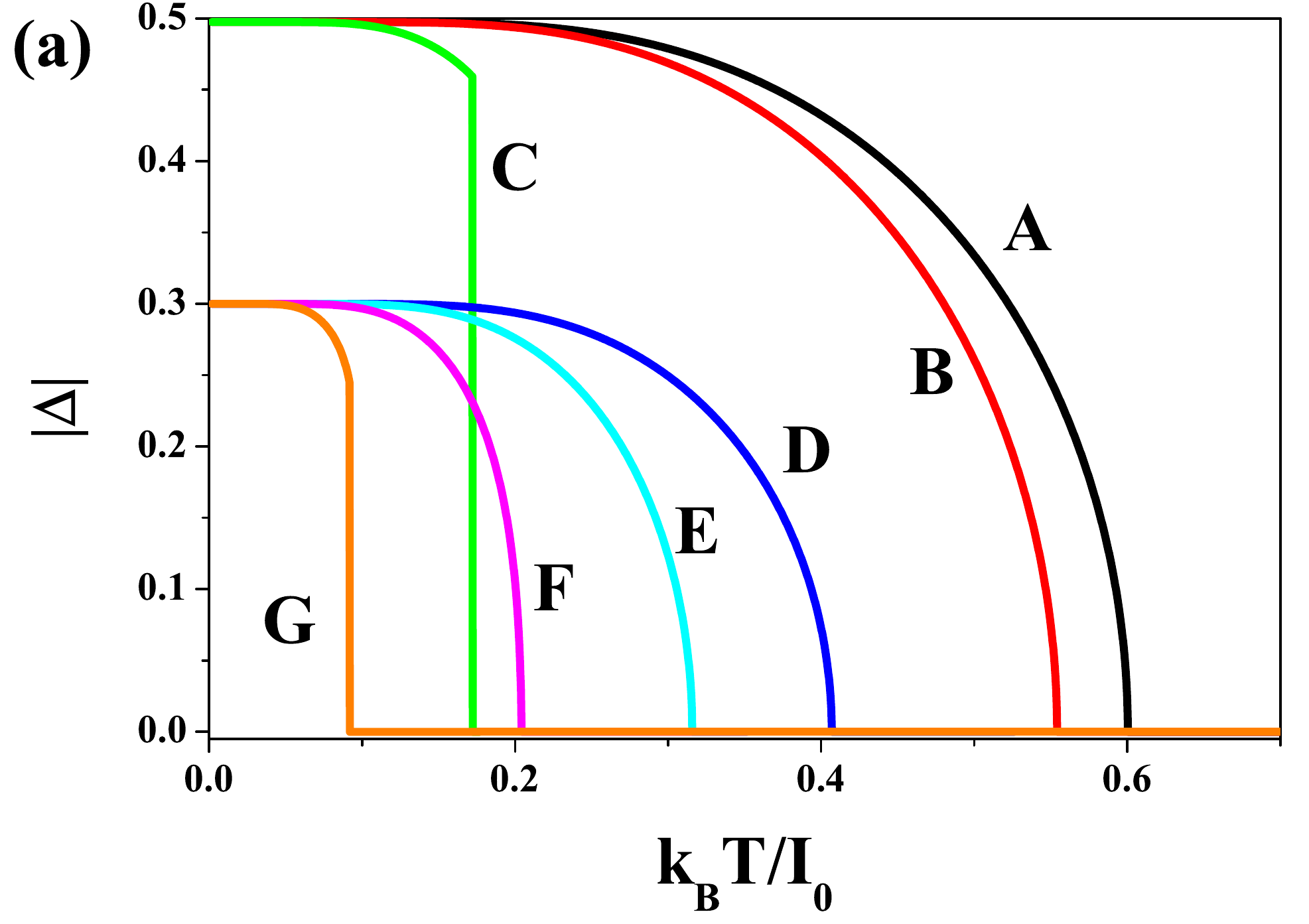}
    \includegraphics[width=0.32\textwidth]{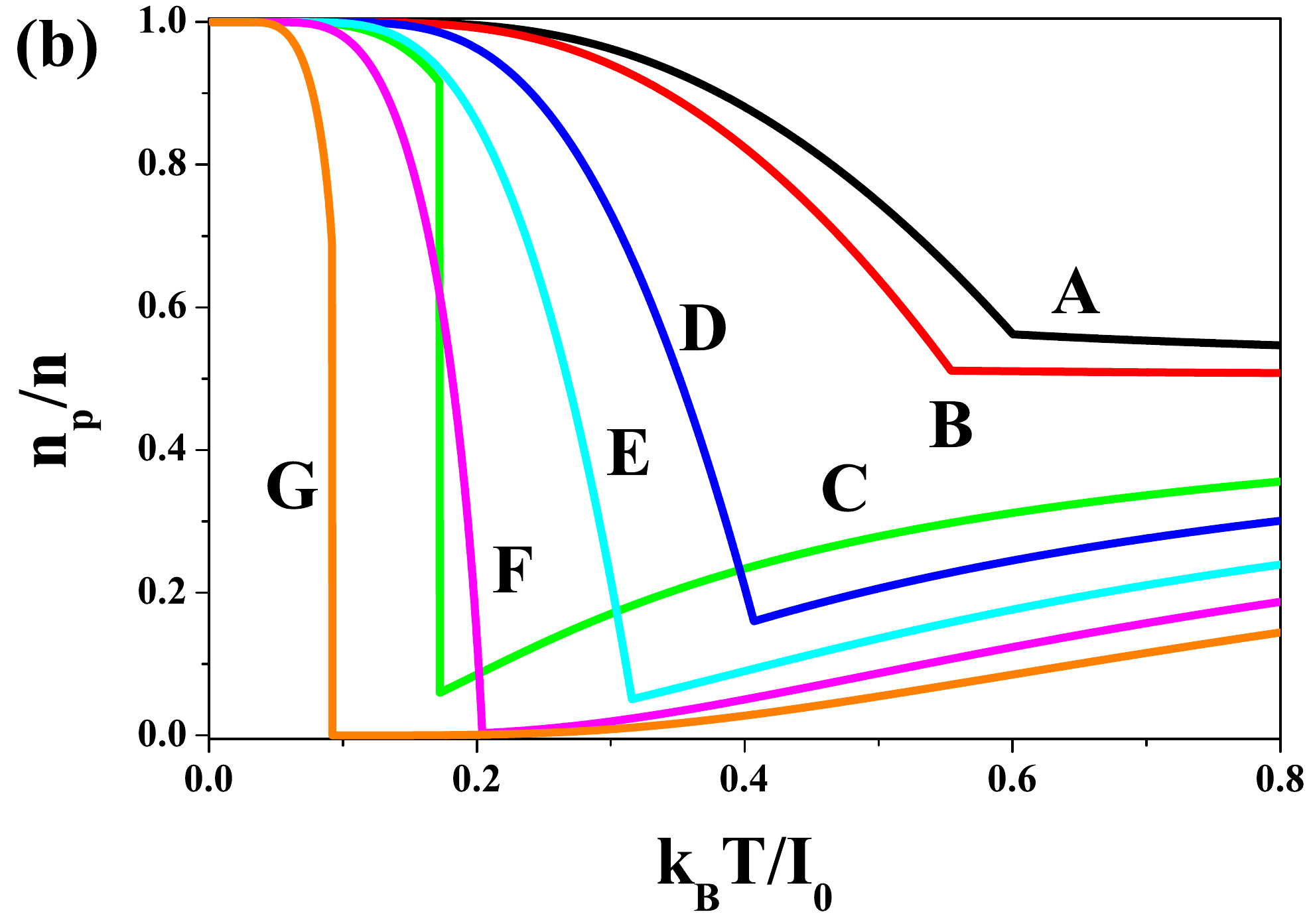}
    \includegraphics[width=0.32\textwidth]{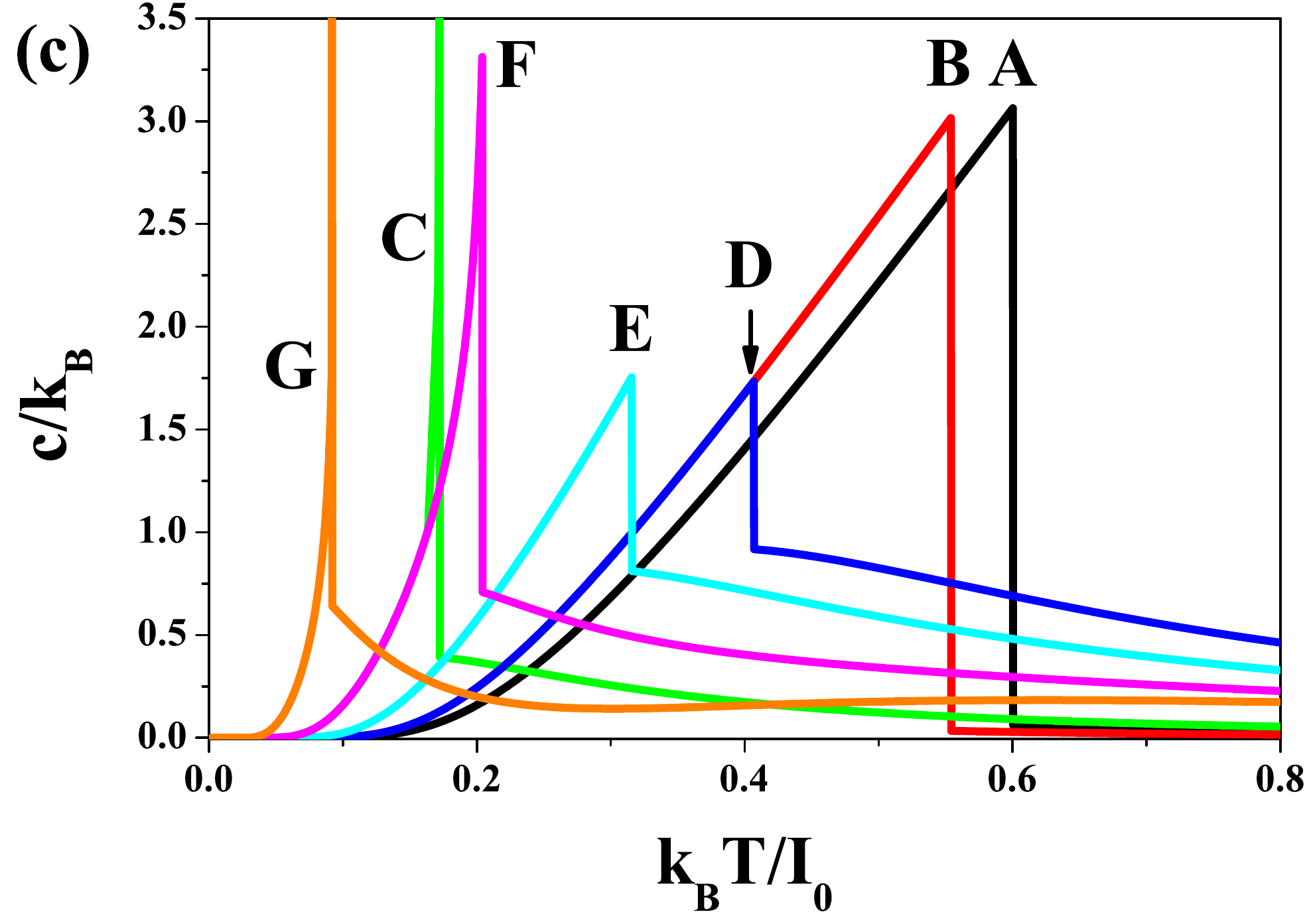}
    \caption{(Color online)
    Temperature dependencies of (a)~the superconducting order parameter $|\Delta|$ (b)~the ratio $n_p/n$ and (c)~the specific heat $c/k_B$ plotted for: \mbox{$\bar{\mu}/I_0=-0.1$} and \mbox{$U/I_0 = -0.5$} (A), \mbox{$U/I_0 = -0.25$} (B), \mbox{$U/I_0 = 0.75$} (C)
    as well as for:
    \mbox{$\bar{\mu}/I_0=-0.8$} and \mbox{$U/I_0 = -0.25$} (D), \mbox{$U/I_0 = 0.25$} (E), \mbox{$U/I_0 = 0.75$} (F) \mbox{$U/I_0 = 1.25$} (G).
    In the panel (c) the lines  C-F and D-B are not distinguishable at low temperatures.}
    \label{rys:HomoProp}
\end{figure*}

For given value of $n$ the $k_BT/I_0$~vs.~$U/I_0$ phase diagrams of the system are shown in figure~\ref{rys:PSkTvsU}. All of the transition temperatures decrease with increasing $U/I_0$.  In the range \mbox{$0<n<1$} the diagrams are rather similar and they consist of three regions (exemplary diagrams are shown in figures~\ref{rys:PSkTvsU}a, b). The regions of SS and NO phases stability are separated by second order boundary or, for larger on-site repulsion, by the region of PS state occurrence. For half-filling (\mbox{$n=1$}, \mbox{$\bar{\mu}=0$}) only homogeneous phases are present on the diagram (figure~\ref{rys:PSkTvsU}c). The $\mathbf{T}$-point, connected with the change of the \mbox{SS--NO} transition order, is located at \mbox{$k_BT/I_0=1/3$} and \mbox{$U/I_0=(2/3)\ln 2$} (for \mbox{$n=1$}).

The possible sequences of transitions with increasing temperatures and the transition orders of them are listed below:
\begin{itemize}
\item[(i)]{SS$\rightarrow$NO: second order for \mbox{$n\neq1$} and second order or first order for \mbox{$n=1$} (and \mbox{$U/I_0<2$}),}
\item[(ii)]{PS$\rightarrow$NO: ''third order'', it can take place only for \mbox{$n\neq1$} (and \mbox{$1<U/I_0<2$}),}
\item[(iii)]{SS$\rightarrow$PS$\rightarrow$NO: both ''third order'', it can take place only for \mbox{$n\neq1$} (and \mbox{$(2/3)\ln 2<U/I_0<2$}).}
\end{itemize}

Notice that increasing temperature usually suppresses heterogeneity in a~system what favours occurrence of homogeneous phases rather than PS states. In the model considered the PS state can develop at higher temperatures than the homogeneous SS phase. This rather unusual behaviour can be connected with the interplay between the pair hopping interaction and the on-site repulsion.


\section{Thermodynamic properties (VA)}\label{sec:properties}

\begin{figure*}
    \centering
    \includegraphics[width=0.32\textwidth]{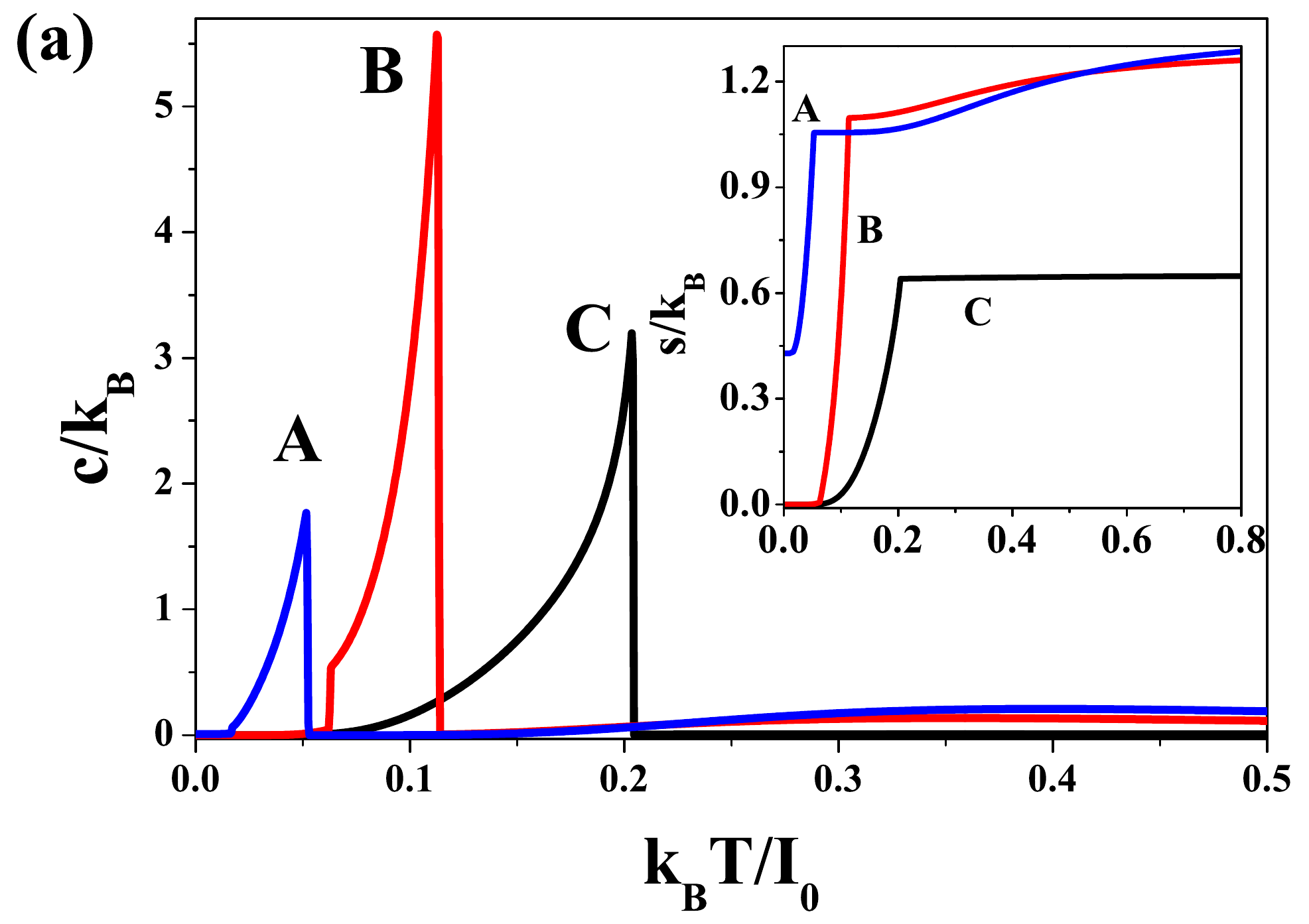}
    \includegraphics[width=0.32\textwidth]{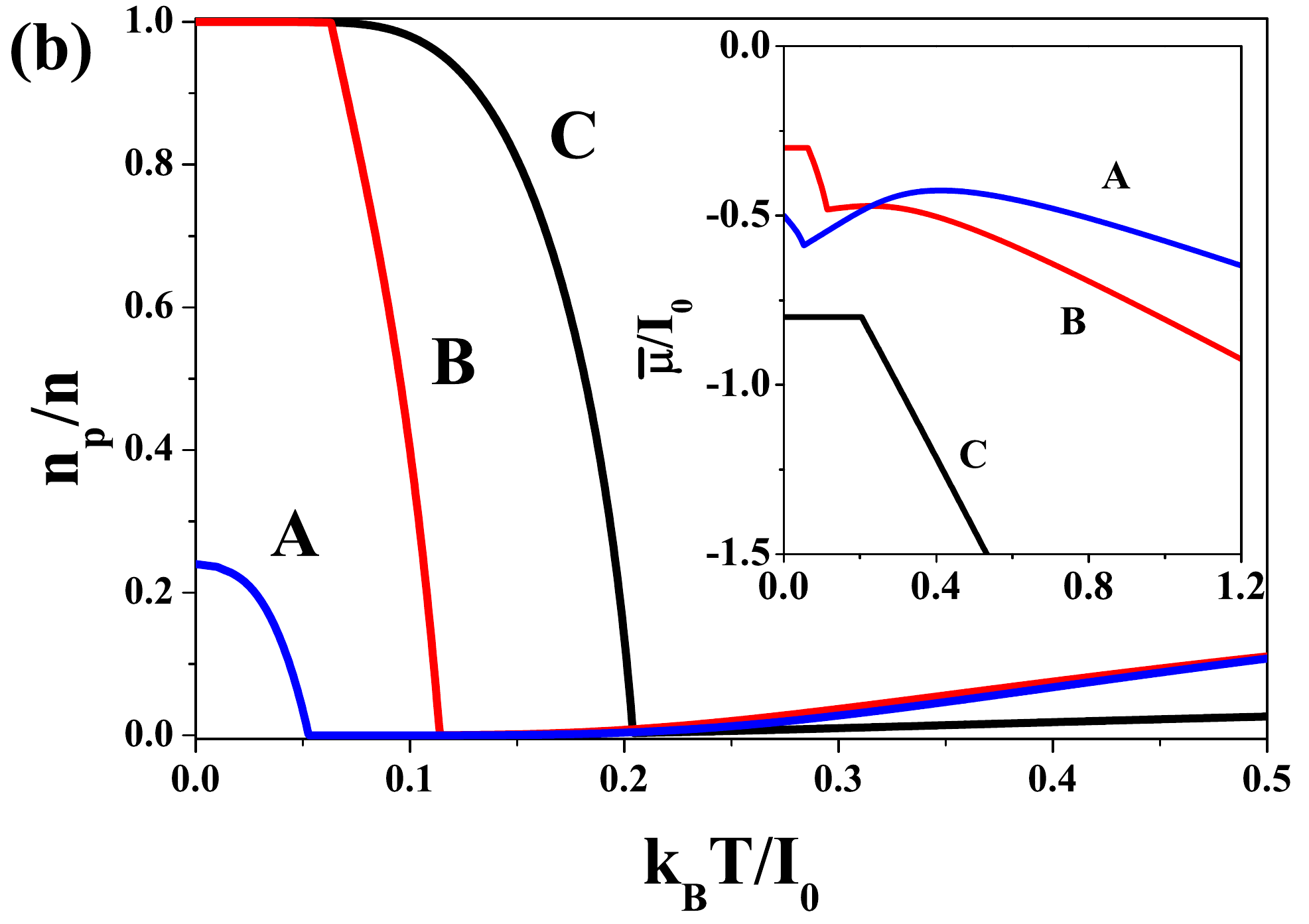}
    \includegraphics[width=0.32\textwidth]{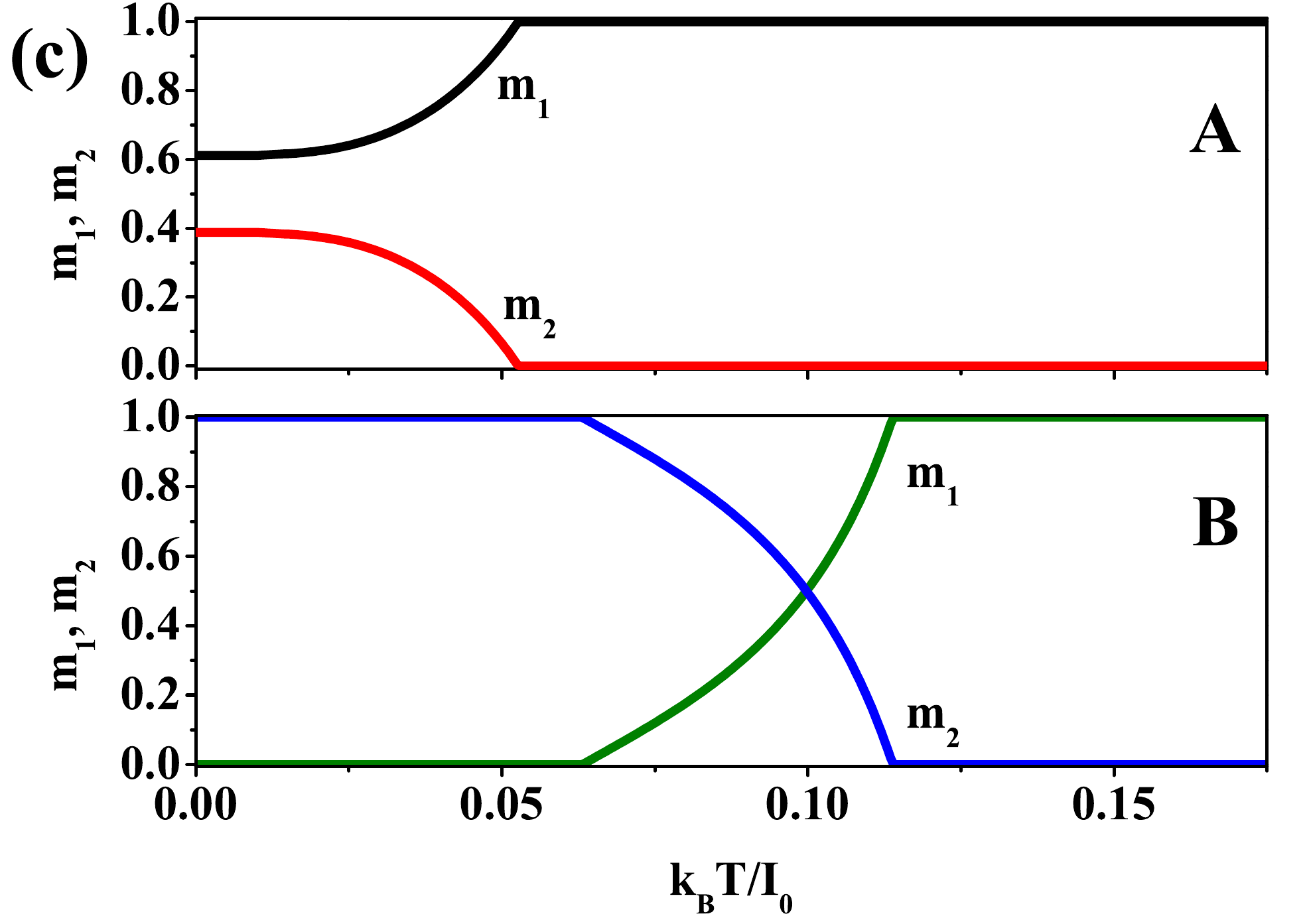}
    \caption{(Color online)
    Temperature dependencies of (a)~the specific heat $c/k_B$, (b)~the ratio $n_p/n$, and (c)~the fractions $m_1$ and $m_2$ of the system  with a charge density $n_+$ and $n_-$, respectively. On the insets there are temperature dependencies of the entropy $s/k_B$~(a) and the chemical potential $\bar{\mu}/I_0$~(b). The characteristics are obtained for: \mbox{$n=0.8$} and \mbox{$U/I_0=1.25$ (A), \mbox{$n=0.7$} and \mbox{$U/I_0=1$} (B), and \mbox{$n=0.2$} and $U/I_0=0.75$} (C).
    }
    \label{rys:SepPropFunkkT}
\end{figure*}
\begin{figure*}
    \centering
    \includegraphics[width=0.32\textwidth]{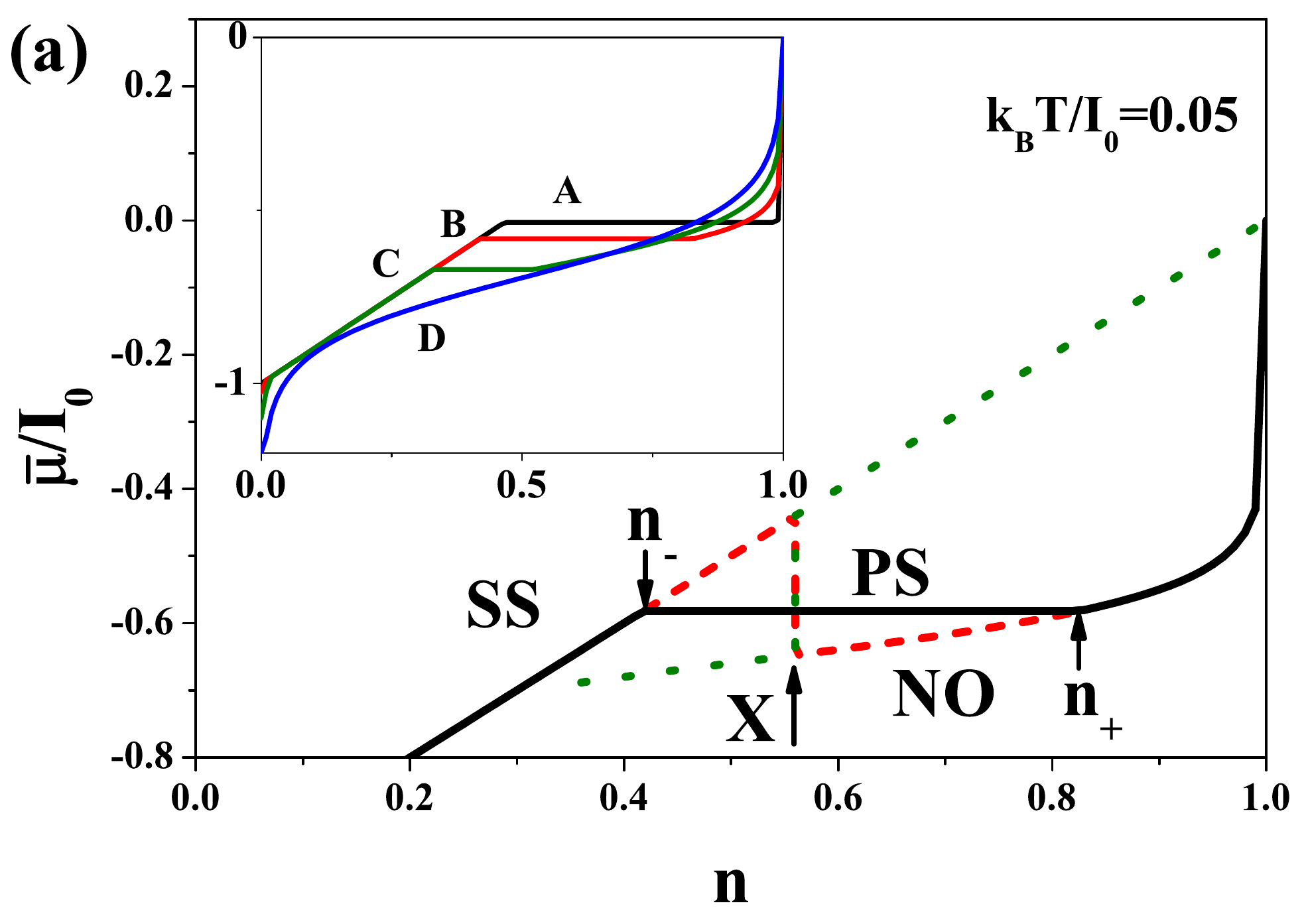}
    \includegraphics[width=0.32\textwidth]{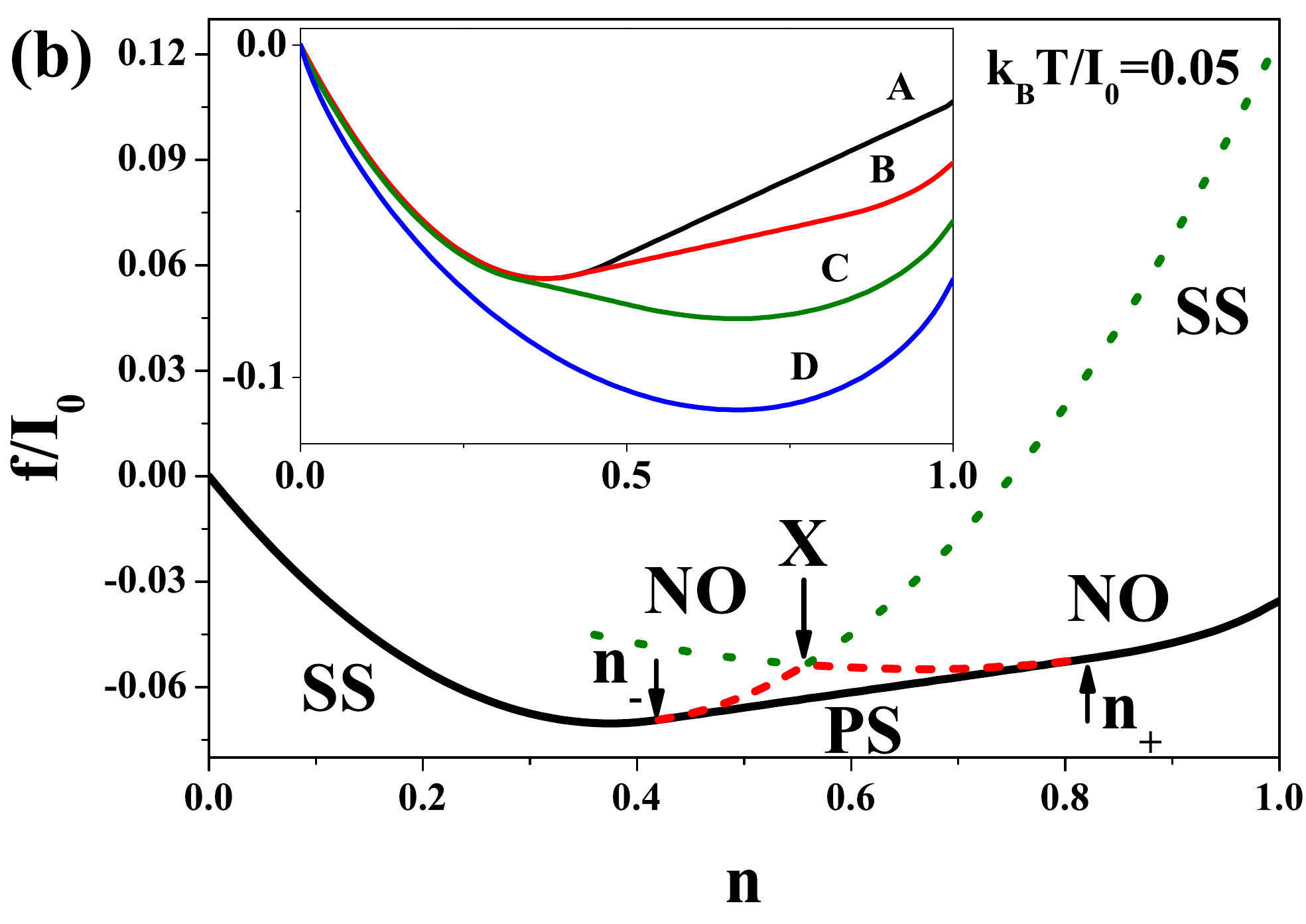}
    \includegraphics[width=0.32\textwidth]{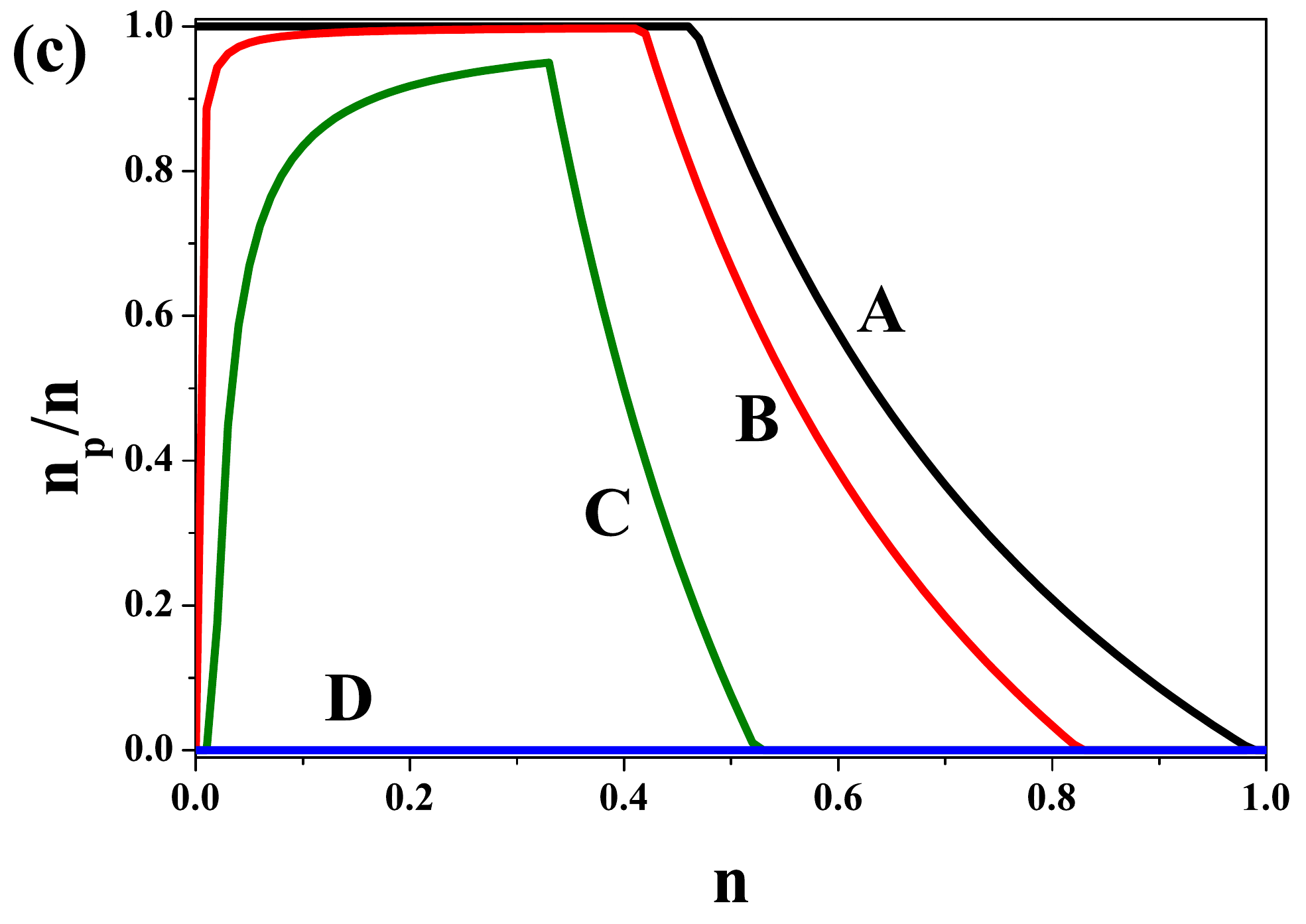}
    \caption{(Color online)
    Concentration  dependencies of (a)~the chemical potential $\bar{\mu}/I_0$, (b)~the free energy $f/I_0$,  and (c)~the ratio $n_p/n$ for \mbox{$U/I_0=1.25$} at fixed temperatures. The dependencies are obtained for: \mbox{$k_BT/I_0=0.025$} (A), \mbox{$k_BT/I_0=0.05$} (B), \mbox{$k_BT/I_0=0.075$} (C), and \mbox{$k_BT/I_0=0.1$} (D). The dashed and dotted lines on panels (a) and (b) correspond to the lowest and highest energy homogeneous metastable phases, respectively. Details in text.}
    \label{rys:SepPropFunkn}
\end{figure*}

In this section we present several representative dependencies  of the thermodynamic characteristics for fixed model parameters (figures~\ref{rys:HomoProp}--\ref{rys:SepPropFunkn}).

In particular for fixed $\bar{\mu}$ (figure~\ref{rys:HomoProp}), one can single out two limiting types of thermodynamic behaviour near transition temperature $T_{SS}$: (i)~the \textit{local pair regime}
and (ii)~the \textit{pair breaking regime}.
In between, there is a~crossover between the two regimes (see e.~g. the plots $n_p/n$ vs. $k_BT/I_0$ in figure~\ref{rys:HomoProp}b).

For large on-site attraction the concentration of locally paired electrons \mbox{$n_p=2D$} exhibits no sharp feature as the temperature is lowered through $T_{SS}$. The number of non-paired electrons at $T_{SS}$ is negligible and the transition is to  the state of dynamically disordered pairs (only for \mbox{$|U|/I_0\gg1$} and \mbox{$U<0$}, the \textit{local pair regime}).

In the second limit, for on-site repulsion \mbox{$U\lesssim 2I_0$}, $n_p$ has a sharp break at $T_{SS}$ and a~substantial fraction of single particles can exists above $T_{SS}$. We call this the \textit{pair breaking regime}. As temperature is lowered, the condensate growths  both from a~condensation  of pre-existing pairs and from
binding and condensation of single particles.
For small binding energies,  if \mbox{$n\ll1$} ($\bar{\mu}\approx-1$), there will be essentially no pre-formed pairs at $T_{SS}$.

Notice that non-zero value of $n_p$ does not imply that local pairs are in coherent state and even significant values of $n_p/n$ are possible in the NO phase. In the limit \mbox{$T\rightarrow+\infty$} $n_p$ increases to \mbox{$n_p/n\rightarrow 0.5$} (each of four states at a~given site can be occupied with equal probability).
The condensate density (which can be approximated as \mbox{$n_0\approx |\Delta|^2$} at least for \mbox{$n\ll1$}, \mbox{$n_0\neq D$}) vanishes for \mbox{$T\geq T_{SS}$},  but the doubly occupied states are still thermally excited above $T_{SS}$ (\mbox{$D\neq0$}).

In figure~\ref{rys:HomoProp}a  the temperature dependencies of the superconducting order parameter $\Delta$ are presented, where one can see clearly the discontinuous change of the order parameter (lines C and G). The other lines correspond to second order transitions.

Finally, let us briefly summarize the behaviour of the specific heat at constant volume \mbox{$c=-T\left(\frac{\partial^2 \omega}{\partial T^2}\right)_{\bar{\mu}}$} (figure~\ref{rys:HomoProp}c). The NO phase is characterized by the relatively broad maximum in $c$  connected with continuous changes in a~short-range electronic ordering (in higher temperatures, not shown in figure~\ref{rys:HomoProp}).  The narrow peak in $c(T)$ is associated with the first order transition, while the $\lambda$-point behaviour is typical for the second order transition.

Let us concentrate now on the thermodynamic properties of the system in the PS state and their changes at the ``third order'' transitions for fixed electron concentration $n$ (figures~\ref{rys:SepPropFunkkT} and~\ref{rys:SepPropFunkn}). As an example, we consider the temperature dependencies of: (i)~the specific heat \mbox{$c=-T\left(\frac{\partial^2 f}{\partial T^2}\right)_{n}$}, (ii)~the entropy \mbox{$s=-\frac{\partial f}{ \partial T}$}, (iii)~the concentration of paired electrons $n_p/n$, (iv)~the chemical potential $\bar{\mu}/I_0$, and (v)~the fractions \mbox{$m_1=(n-n_-)/(n_+-n_-)$} and \mbox{$m_2=1-m_1$} (sizes of domains) with a~charge density $n_+$ and $n_-$ (figure~\ref{rys:SepPropFunkkT}) as well as the concentration dependencies of: (a)~$\bar{\mu}/I_0$, (b)~$f/I_0$ and (c)~$n_p/n$ (figure~\ref{rys:SepPropFunkn}).

Thermodynamic characteristics  for three different possible sequences of transitions with increasing temperature are shown in figure~\ref{rys:SepPropFunkkT}. The label (A) corresponds to the case of \mbox{PS$\rightarrow$NO} transition. At such a~transition the specific heat $c$ exhibits a~finite jump at transition temperature, what is associated with the continuity of the entropy $s$ (cf. figure~\ref{rys:SepPropFunkkT}a). The lines labelled by (B) correspond to \mbox{SS$\rightarrow$PS$\rightarrow$NO} case. At the \mbox{SS-PS} transition  there is also a~finite jump in $c$, however the $c$ in the PS state is larger than that in the SS phase, contrary to the \mbox{PS-NO} transition, where the jump has an opposite sign.
Finally, the lines (C) correspond to the case of second order SS$\rightarrow$NO transition.
The concentration ratio of locally paired electrons $n_p$ is presented in figure~\ref{rys:SepPropFunkkT}b.
On the inset of figure~\ref{rys:SepPropFunkkT}b the chemical potential  is shown as a function of temperature.  One can see that $\bar{\mu}$ in the SS phase is independent of $T$ (cf.~(\ref{row:miSS})).
In figure~\ref{rys:SepPropFunkkT}c there are shown dependencies of the fractions of the system $m_1$ and $m_2$ with charge density $n_+$ and $n_-$, respectively. For instance, in the case (A) the SS domain with lower concentration ($n_-$) vanishes at ``third order'' transition.

The chemical potential $\bar{\mu}/I_0$ and the free energy $f/I_0$ as a function of concentration for \mbox{$U/I_0=1.25$} and \mbox{$k_BT/I_0 = 0.05$} are presented in figure~\ref{rys:SepPropFunkn}a and figure~\ref{rys:SepPropFunkn}b, respectively.
For \mbox{$n<n_-$} the SS phase is stable and its free energy is the lowest one. In the range \mbox{$n_-<n<n_+$} the PS state has the lowest energy (solid line).
The solutions of \mbox{(\ref{row:MFA1})--(\ref{row:MFA2})} corresponding to the lowest energy ones (the homogeneous phases with the lowest energy)  are thermodynamically stable i.~e. $\partial \bar{\mu}/\partial n>0$ in their existence ranges (denoted by dashed lines in figures~\ref{rys:SepPropFunkn}a,~b). At point labelled as $\mathbf{X}$ the transition between metastable phases occurs, eg. for \mbox{$n<n_X$} the free energy $f_{SS}$ of  the SS phase is lower than the energy $f_{NO}$ of the NO phase, whereas \mbox{$f_{SS}>f_{NO}$} for \mbox{$n>n_X$}.
The dotted lines in figures~\ref{rys:SepPropFunkn}a,~b correspond to the homogeneous phases with the highest energy. One should notice that in the described case for \mbox{$U/I_0=1.25$} and \mbox{$k_BT/I_0=0.05$} in the whole range of the PS state occurrence two homogeneous phases are metastable, whereas in the ranges of the homogeneous phases stability there is one metastable phase at the most.
The linear dependence \mbox{$\bar{\mu}/I_0=n-1$} is characteristic for the SS phase, whereas in the PS state $\bar{\mu}/I_0$ is independent of $n$. This is consistent with square dependence of $f_{SS}$ and linear dependence of $f_{PS}$ as a function of $n$.

The anomalous decrease of the chemical potential with doping in the lowest energy metastable homogeneous phases near $\mathbf{X}$-point (i.~e. an abrupt drop at $\mathbf{X}$-point) is a~signal that it is thermodynamically more convenient to phase separate into two subsystems with the different electron densities $n_-$ and $n_+$ (figure~\ref{rys:SepPropFunkn}a). The density range for which the PS occurs can be readily found by the Maxwell's construction described in section~\ref{sec:method}.

The $n$-dependencies of the chemical potential $\bar{\mu}/I_0$, the free energy $f/I_0$  and the concentration of paired electrons \mbox{$n_p/n=2D/n$} are shown on the insets of figure~\ref{rys:SepPropFunkn}a and figure~\ref{rys:SepPropFunkn}b and in figure~\ref{rys:SepPropFunkn}c, respectively, for \mbox{$U/I_0=1.25$} at several fixed temperatures. They are obtained for: \mbox{$k_BT/I_0=0.025$} (A), \mbox{$k_BT/I_0=0.05$} (B), \mbox{$k_BT/I_0=0.075$} (C), and \mbox{$k_BT/I_0=0.1$} (D).

At ``third order'' transitions (SS-PS and PS-NO) the thermodynamic parameters such as $\bar{\mu}$, $n_p/n$, $s$, $f$ are continuous, whereas the $c$ exhibits a~finite jump as at the second order transitions. However, one should notice that the order parameter for ``third order'' transitions is the concentration difference \mbox{$n_+-n_-$} (not $\Delta$, which is the order parameter in one domain) and its change is discontinuous at transition temperature. Such transitions are present if the system is consider for fixed $n$ and they are associated with first order transitions at fixed $\bar{\mu}$.


\section{Final remarks}\label{sec:conclusions}

We have studied a~simple model of a~superconductor with very short coherence length (i.~e.~with the pair size being of the order of the radius of an effective lattice site) and considered the situation where the single particle mobility is much smaller than the pair mobility and can be neglected.
We have evaluated the phase diagrams and thermodynamic characteristics of the model at fixed $\mu$ and at fixed $n$ and determined the ranges of the PS occurrence in the system.

One should stress that  within the VA the on-site $U$ term is treated exactly. Thus, the major conclusions of our paper concerning the evolution of the properties of the system with $U$ are reliable for arbitrary $U$.
The mean-field used for the intersite term is best justified if the $I_{ij}$ interactions are long-ranged or if the number of nearest neighbours is relatively large.
The derived VA results are exact in the limit of infinite dimensions \mbox{$d\rightarrow+\infty$}, where the MFA treatment of the intersite interaction $I$ term becomes the rigorous one. Moreover, the VA yields exact results (in the thermodynamic limit) for $I_{ij}$ of infinite range (\mbox{$I_{ij}=(1/N)I$} for any $(i,j)$) regardless of the dimensionality of the system~\cite{HB1977}.

The properties of the system are strongly dependent on  the ratio of $U/I$, which is related to the relative values of the pair binding energy $E_b$  and the effective pair mobility $t_p$. For the case of a~single pair in a~system, i.~e. for \mbox{$n\rightarrow0$}, we have \mbox{$E_b=-U+2I_0$}, \mbox{$t_p=2I$} (the pair effective mass is \mbox{$m^*_p=1/(2t_p)=1/(4I)$})  \cite{RP1993} and increasing  $n$ can modify $E_b$. There are two well defined limits of the model: (i)~the \textit{local pair regime} and (ii)~the \textit{pair breaking limit}. In between there is a~crossover between the two regimes for fixed $\mu$.

The \textit{local pair limit} is realized for large pair binding energy (i.~e. for \mbox{$U<0$}). In such a~case, the two interactions: the on-site interaction $U$ and the intersite pair hopping $I$ cooperate and transition temperature is determined by pair mobility (i.~e. by $I$). The \mbox{SS-NO} transition with increasing temperature is second order and is to the NO phase being a~state of dynamically disordered local pairs.

The opposite regime i.~e. the \textit{pair breaking limit} is realized for substantial values of  \mbox{$U>0$}.
Repulsive $U$ destabilizes on-site pairing and competes with $I$. As we have shown above, the increasing $U$: (i)~for fixed $\mu$ it changes first the nature of phase transitions from a~continuous to a discontinuous type and then it suppresses superconductivity for \mbox{$|\bar{\mu}|/I_0=|n-1|\rightarrow1$}, (ii)~it stabilizes the phase separation state (\mbox{SS--NO}) in definite ranges of $n$ and temperature. This behaviour is associated with the presence of the tricritical point on the phase diagrams.
With increasing $U/I_0$ the maximum of the transition temperature moves towards lower concentrations (for \mbox{$n<1$}). In definite range of $U/I_0$ and $n$ the PS state \mbox{SS--NO} is stable. For \mbox{$1<U/I_0<2$} the PS state extends from the ground state, whereas for \mbox{$(2/3)\ln 2<U/I_0\leq1$} it is stable only at finite temperatures.

Finally, for large $U$ (\mbox{$U/I_0>2$}), the system remains in a~normal state at all temperatures for any chemical potential or concentration.

Comparing the VA results at \mbox{$T=0$} with the other ones (exact ones for \mbox{$d=1$} chain, RPA ones for \mbox{$d=2$} square lattice and \mbox{$d=3$} cubic lattices),  one can see that the general structure of the diagrams is similar. It implies that the VA can give qualitatively reasonable results also for lattices of finite dimensionality.
However, one should stress that in finite dimensions due to quantum fluctuations connected with $I_{ij}$ interactions the regions of the homogeneous SS phase occurrence are extended in comparison with the VA results. Moreover, in a~linear chain only the short-range order occurs at \mbox{$T=0$} in contrary to the VA and RPA (SWA) in \mbox{$d\geq 2$}, where the long-range order is present.

One should emphasize that for \mbox{$d<+\infty$} the VA overestimates transition temperatures and gives incorrect concentration dependence of them (\mbox{$k_BT_{SS}\sim -1/ \ln(n/2)$} for \mbox{$n\ll1$}, cf.~(\ref{row:2ndorderTSSNO})). For \mbox{$d=3$} ideal Bose gas with $k^2$ spectrum (the \mbox{$U\ll0$} and \mbox{$n\ll1$} limit of the model considered) the $n$-dependence of the \mbox{SS--NO} temperature is  \mbox{$T_{SS}\sim n^{2/3}$}, what can be obtained, for example in the RPA, even for finite values of $U/I_0$~\cite{MR1992,M0000}. The RPA predicts also the existence of tricritical point in finite temperatures~\cite{M0000} in \mbox{$d=3$}, what confirms that the VA can give qualitatively reasonable results also for real lattices in finite temperatures. The RPA results for \mbox{$d=2$} at \mbox{$T>0$} \cite{MR1992,M0000} are in agreement with the Mermin-Wagner theorem \cite{MW1966}, which proofs the absence of  the long-range order for \mbox{$d\leq2$} at \mbox{$T>0$}. In \mbox{$d=2$} one has
the Kosterlitz-Thouless-type transition with increasing temperature.

The model considered can describe the systems in which the single electron mobility (single electron hopping) is much smaller than the pair mobility (kinetic energy of pairs) and it can be neglected in first approximation. Let us comment on the finite electron bandwidth effects.
The presence of the hopping term \mbox{$\sum_{i,j,\sigma}t_{ij}\hat{c}^+_{i\sigma}\hat{c}_{j\sigma}$}
breaks a~symmetry between \mbox{$I>0$} (favouring SS) and \mbox{$I<0$} (favouring $\eta$S) cases.
The phase diagrams of the Penson--Kolb--Hubbard model are more complicated than those obtained for Hamiltonian (\ref{row:ham1}) \cite{HD1993,RB1999,JKS2001,CR2001,DM2000,Z2005,MM2004,MC1996}, even in the ground state.
We can suppose that small but finite single electron hopping $t_{ij}$ will not qualitatively alter the phase diagrams, at least for the case \mbox{$k_BT_{SS}>\sum_jt^2_{ij}/U$}. The main effect of $t_{ij}$ (for \mbox{$|U|\gg t_{ij}$}, \mbox{$U<0$})  is a~renormalization of the pair hopping term \mbox{$I_{ij}\rightarrow I_{ij}+2t^2_{ij}/|U|$} and an introduction of an effective intersite density-density repulsion \mbox{$\sim t^2_{ij}/|U|$}. For \mbox{$U<0$} and \mbox{$I<0$} the charge density wave state can also occur \cite{RB1999,JKS2001}. For \mbox{$U>0$} and both signs of $I$ the $t_{ij}$ term generates magnetic correlations
competing with superconducting ones and its effects essentially modify the phase diagrams and the properties of normal state (particularly for larger values of $t_{ij}$). In such a~case it is necessary to consider also various  magnetic orderings. Moreover, several phase separation states involving superconducting, charge and (or) magnetic orderings could also be stable  for \mbox{$n\neq 1$}.
For larger values of $t_{ij}$ the studies of GS phase diagrams of the half-filled PKH model for $d$-dimensional lattices have been determined by means of the (broken symmetry) Hartree-Fock approximation  (HFA) and by the slave-boson mean-field method (SBMFA) in \cite{RB1999}.
For \mbox{$d=1$} the diagrams supported also by the results of the continuum-limit field theory and exact diagonalisation studies \cite{JKS2001} are shown to consist of at least nine different phases including superconducting states, bond located antiferromagnetic and charge-density-wave states as well as mixed phases with coexisting site and bond orderings. The stability range of the bond-type orderings shrinks with increasing lattice dimensionality and for \mbox{$d\rightarrow + \infty$} the phase diagrams involve exclusively site-located orderings.
One should notice that the phase boundary at \mbox{$T=0$} and \mbox{$n=1$} determined by the HFA and SBMFA (\mbox{$t_{ij}\neq0$}) are in agreement with our VA result \mbox{$U/I_0=1$}, even for relatively large $t_{ij}$ (especially for \mbox{$I>0$}) \cite{RB1999}. Obviously, for \mbox{$t_{ij}\neq0$} the transition is to the antiferromagnetic Mott phase.

\begin{figure}
    \centering
    \includegraphics[width=0.49\textwidth]{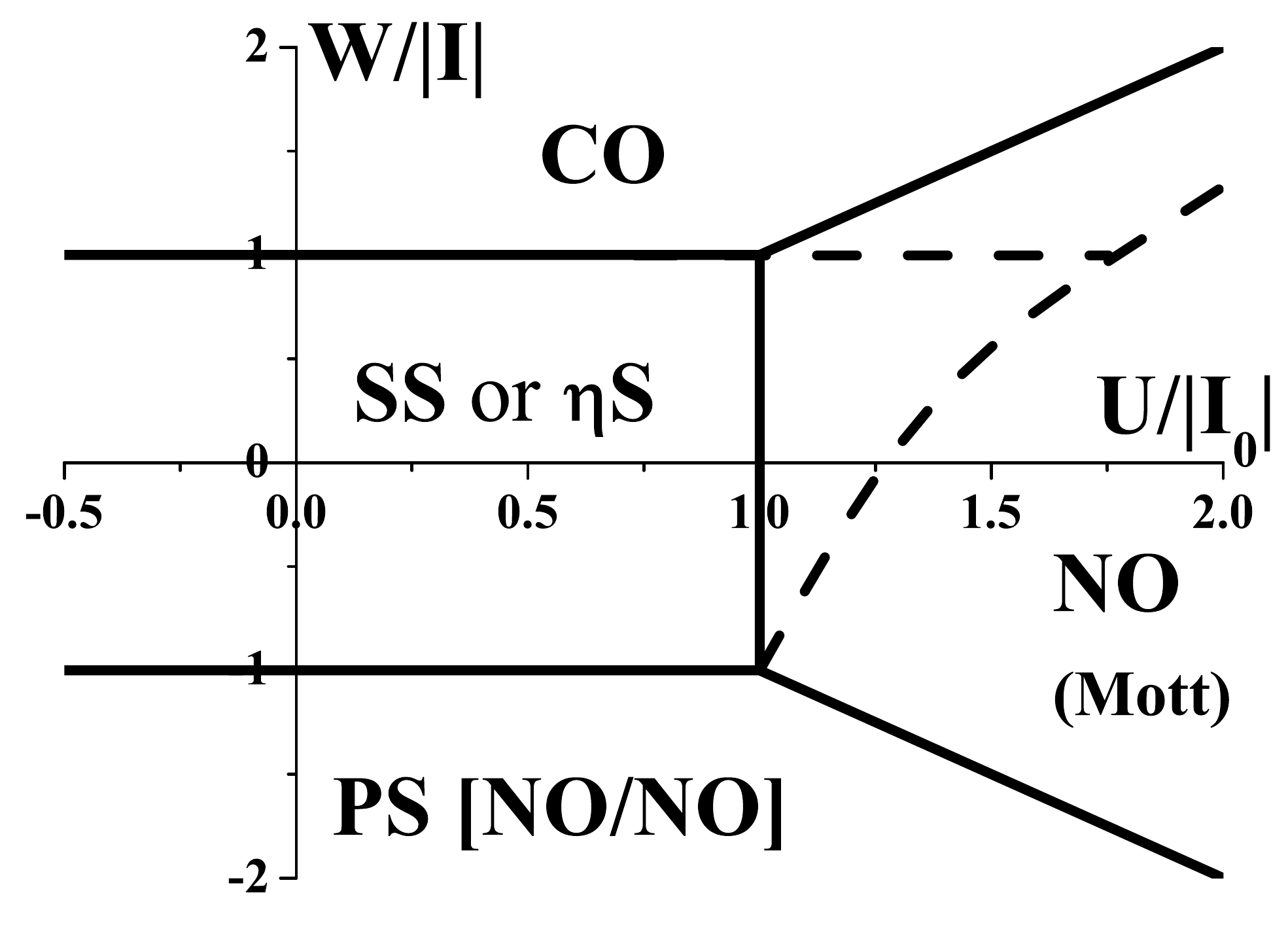}
    \caption{Ground state phase diagram in presence of density-density interaction $W$ between nearest neighbours for half-filling obtained within VA (solid lines). The dashed lines denote the rigorous boundaries for 1D-chain. All transitions between homogeneous phases are discontinuous. \mbox{$I_0=zI$}.}
    \label{rys:UWI}
\end{figure}

The intersite density-density interactions $W_{ij}$ can introduce mixed phases and (or) phase separations between charge orderings and superconductivity \cite{RP1996}. The ground state phase diagram in presence of interaction $W$ between nearest neighbours (added a~term \mbox{$(W/2)\sum_{\langle i,j \rangle} \hat{n}_i \hat{n}_j $} to Hamiltonian (\ref{row:ham1})) calculated for half-filling is presented in figure~\ref{rys:UWI} (cf. with \cite{JS2002}, where the case \mbox{$t_{ij}\neq0$} has been analyzed).
For sufficiently strong intersite repulsion \mbox{$W>0$} the homogeneous charge ordered (CO) phase occurs (with \mbox{$n_Q = (1/N)\sum_{i} \langle \hat{n}_i \rangle \exp{(i\vec{Q}\cdot\vec{R}_{i})}$}), whereas intersite attraction \mbox{$W<0$} introduce the PS state  in which two NO phases coexists (with \mbox{$n_+=2$} and \mbox{$n_-=0$}, i.~e. electron droplets, PS[NO/NO]). For $d=1$ the regions of the SS (or $\eta$S, depending on the sign of $I$) and CO phases occurrence are extended,whereas a~region of the NO (Mott) phase is reduced.
The superconducting order (in $d=1$), similarly as in section~\ref{sec:GSexact}, is a~short-range order whereas the CO is a~long-range order (cf. also \cite{MM2008}).

The electron concentration $n$ and chemical potential $\mu$ are (thermodynamically) conjugated  variables in the bulk systems. However, one can fit the concentration rather than the potential in a~controlled way experimentally. In such a~case $\mu$ is a~dependent internal parameter, which is  determined by the temperature, the value of $n$, and other model parameters (cf. (\ref{row:condn1})). Thus the obtained phase diagrams as a~function of the concentration are quite important  because in real systems $n$ can vary in a~large range.

Although our model is (in several aspects) oversimplified, it can be useful in qualitative analysis of experimental data for real narrow-band materials and it can be used for better understanding of various systems mentioned below and in section~\ref{sec:intro}. In particular, our results predict the existence of the phase separation (\mbox{SS--NO}) near the Mott state and describe its possible evolution and phase transitions with increasing temperature and a~change of $n$ ($\bar{\mu}$).

Of course the PS instability is specific to the short-range nature of the interactions in the model. When the (unscreened) long-range Coulomb interactions are included the large-scale PS of charged particles is prevented and only a~frustrated PS can occur (mesoscale, nanoscale) with the formation of various possible textures~\cite{EK1993,JCC2001,YY2004}.

The electron phase separation involving SS is shown experimentally in several systems.
In particular, for special cases of  La$_2$CuO$_{4+\delta}$ and La$_{2-x}$Sr$_x$CuO$_{4+\delta}$, muon and superconducting quantum interference measurements  suggest that the electron inhomogeneities move beyond local variations to form fully phase separated regions \cite{UAC2009,SFG2002,MWB2006}.
Organic compounds also exhibit the superconductor-insulator phase separations as a~result of the external pressure (e.~g. quasi-one dimensional (TMTSF)$_2$PF$_6$ \cite{KPK2004}, (TMTSF)$_2$ReO$_4$ \cite{CSP2008}) and fast cooling rate through the glass-like structure transition (e.~g. $\kappa$-(ET)$_2$Cu[N(CO)$_2$]Br \cite{TCS2008}).
Finally we mention the family of iron-pnictides, e.~g. in Ba$_{1-x}$K$_x$Fe$_{2-y}$Se$_2$, where mesoscopic PS between SS and insulating (magnetic) phases have been observed \cite{PIN2009,RPC2011}.

The detailed analysis of RPA solutions will be explored further in subsequent publication~\cite{M0000}.
It is of interest to analyze the effects of external magnetic field as well as  the impact of density-density \cite{MM2008,RMC1984} and magnetic \cite{KKR2010} interactions on the phase diagrams of model (\ref{row:ham1}).


\begin{acknowledgments}
The authors wish to thank T.~Kostyrko and G.~Paw\l{}owski for helpful discussions and a careful reading
of the manuscript.
The work (K.~K.) has been financed by National Science Center (NCN, Poland) as a research project in years 2011-2013, grant No. DEC-2011/01/N/ST3/00413.
K.~K.  would like to thank the European Commission and Ministry of Science and Higher Education (Poland) for the partial financial support from European Social Fund -- Operational Programme ``Human Capital'' -- POKL.04.01.01-00-133/09-00 -- ``\textit{Proinnowacyjne kszta\l{}cenie, kompetentna kadra, absolwenci przysz\l{}o\'sci}''.
\end{acknowledgments}


\appendix


\section{Site-dependent self-consistent VA equations}\label{app:sitedependentequations}

Within the VA  the on-site interaction term is treated exactly and the intersite interactions are decoupled within the MFA (site-dependent):
\begin{equation}
\hat{\rho}^+_{i}\hat{\rho}^-_{j} \rightarrow \left\langle \hat{\rho}^+_{i} \right\rangle \hat{\rho}^-_{j} + \left\langle \hat{\rho}^-_{j}\right\rangle \hat{\rho}^+_{i} - \left\langle \hat{\rho}^+_{i}\right\rangle \left\langle \hat{\rho}^-_{j}\right\rangle.
\end{equation}
A~variational Hamiltonian has the following form:
\begin{eqnarray}
\hat{H}_{0}& = &\sum_i{\left[U\hat{n}_{i\uparrow}\hat{n}_{i\downarrow}-\mu \hat{n}_i -2\chi^*_i\hat{\rho}^-_i -2\chi_i\hat{\rho}^+_i \right. }+ \nonumber \\
& + & \left. \chi^*_i\Delta_i + \chi_i\Delta_i^* \right],
\end{eqnarray}
where \mbox{$\chi_i=\sum_{j\neq i}I_{ij}\Delta_j$}, \mbox{$\Delta_i^*=\langle\hat{\rho}^+_i\rangle$} and \mbox{$n_i=\langle\hat{n}_i\rangle$}.
$\hat{H}_{0}$ can be diagonalized easily
and a~general expression for the grand potential $\Omega$ in the grand canonical ensemble in the VA is
\begin{equation*}
\Omega= -\frac{1}{\beta}\ln\left\{\textrm{Tr}\left[\exp(-\beta\hat{H_{0}})\right]\right\},
\end{equation*}
where \mbox{$\beta=1/(k_{B}T)$}. The average value of operator $\hat{A}$ is defined as
\begin{equation*}
\langle\hat{A}\rangle = \frac{\textrm{Tr}\left[\exp(-\beta\hat{H}_{0})\hat{A}\right]}{\textrm{Tr}\left[\exp(-\beta\hat{H}_{0})\right]}.
\end{equation*}
$\textrm{Tr} \hat{B}$ means a~trace of operator $\hat{B}$ calculated in the Fock space.

The explicit formula for the grand potential obtained in the VA has the following form
\begin{equation}\label{row:sitedependentgrandpotential}
\Omega = \sum_i\Omega_i = \sum_{i}{\left\{\chi^*_i\Delta_i + \chi_i\Delta^*_i + \frac{U}{2} - \mu  - \frac{1}{\beta}\ln{(2Z_i)}\right\}},
\end{equation}
where
\begin{equation*}
Z_i = \exp{(\beta U/2)} + \cosh{\left( \beta \sqrt{\bar{\mu}^2+4|\chi_i|^2}\right)},
\end{equation*}
and \mbox{$\bar{\mu} = \mu - U/2$}.
The expression for the average number of electrons at $i$-site is given by
\begin{equation}\label{row:nonsite}
n_i - 1= \langle \hat{n}_i\rangle - 1 = \frac{\bar{\mu}\sinh ( \beta \sqrt{\bar{\mu}^2 + 4|\chi_i|^2})}{Z_i\sqrt{\bar{\mu}^2+4|\chi_i|^2}}
\end{equation}
and the average of the charge exchange operator at $i$-site is derived as
\begin{equation}\label{row:rhoonsite}
\Delta^*_i= \langle \hat{\rho}^+_i \rangle = \frac{\sinh (\beta \sqrt{\bar{\mu}^2 + 4|\chi_i|^2})}{Z_i\sqrt{\bar{\mu}^2+4|\chi_i|^2}}\chi^*_i
\end{equation}
so one has a~set of \mbox{$2N$} self-consistent equations to solve consisting of: (i) $N$ equations in form of (\ref{row:nonsite}) and (ii) $N$ equations in form of (\ref{row:rhoonsite}) (two equations for each site from $N$ sites).

The free energy is obtained as \mbox{$F=\Omega + \mu \langle \hat{N}_e \rangle $}, where  \mbox{$\hat{N_{e}}=\sum_{i}{\hat{n}_{i}}$},
\mbox{$\langle\hat{N_{e}}\rangle=nN$} is the number of electrons in the system ($n$ is defined by (\ref{row:condn1})).

The double occupancy $D_i$ of the site $i$ is determined by the following equation:
\begin{equation}\label{row:Doccupsite}
D_i = \left\langle \hat{n}_{i\uparrow} \hat{n}_{i\downarrow}\right\rangle = \frac{n_i}{2}\left[ 1 - \frac{1}{n_i}\frac{\exp{(\beta U/2)}}{Z_i} \right]=\frac{\partial \Omega_i}{\partial U},
\end{equation}
where $n_i$ is defined by (\ref{row:nonsite}).

The solutions of the set \mbox{(\ref{row:nonsite})--(\ref{row:rhoonsite})} can correspond to a~minimum, a~maximum or a~point of inflection of the grand potential (\ref{row:sitedependentgrandpotential}). In other words, the solutions of the set \mbox{(\ref{row:nonsite})--(\ref{row:rhoonsite})} fulfil the necessary condition for a~minimum of $\Omega$ with respect to \mbox{$\{\Delta_i\}_{i=1}^{N}$}. To find the solutions corresponding to stable (or metastable) states of the system, one should find a~minimum of \mbox{$\Omega$} with respect to all $\Delta_i$.

One can show easily that in the case of neglected spatial variations of the superconducting order parameter (i.~e.~\mbox{$\Delta=\Delta_i$} for every site $i$) \mbox{Eqs.~(\ref{row:sitedependentgrandpotential})--(\ref{row:rhoonsite})} reduce to Eqs.~(\ref{row:grandpotential}) and \mbox{(\ref{row:MFA1})--(\ref{row:MFA2})} obtained in section~\ref{sec:method}.


\section{The particle-hole transformation}\label{app:sym}

Let us introduce the particle-hole transformation \mbox{$J=J_{\uparrow}J_{\downarrow}$} on the alternate lattice defined as
\begin{equation*}
J:\quad
\begin{array}{rl}
\hat{c}^+_{i\sigma} & \Rightarrow (-1)^{\lambda_i}\hat{c}_{i\sigma}, \\
\hat{c}_{i\sigma} & \Rightarrow (-1)^{\lambda_i}\hat{c}^+_{i\sigma},
\end{array}
\quad (\textrm{for} \quad \sigma=\uparrow,\downarrow),
\end{equation*}
where \mbox{$\lambda_i=0$} if \mbox{$i\in A$} and \mbox{$\lambda_i=1$} if \mbox{$i\in B$} ($A$, $B$ label two sublattices in the alternating lattice).
Then the operators appearing in Hamiltonian (\ref{row:ham1}) are transformed to
\begin{equation}\label{row:tranformationUminus}
\hat{n}_{i\sigma}  \Rightarrow  1 - \hat{n}_{i\sigma}, \quad
\hat{n}_i  \Rightarrow  2 - \hat{n}_i, \quad
\hat{\rho}^+_i \Rightarrow -\hat{\rho}^-_i
\end{equation}
whereas the thermodynamic parameters and averages: chemical potential, free energy, specific heat, superconducting-order parameter and double occupancy transform as follows
\begin{eqnarray}
\label{row:mutransf}\mu(2-n) &=& - \mu(n) + U, \quad(\ \bar{\mu}(2-n) = -\bar{\mu}(n) \ ) \quad\\
f(2-n) &= &f(n) + U(1-n),\ \ \\
c(2-n)& = &c(n),\\
\Delta(2-n) &=& -\Delta^*(n),\\
D(2-n) &=&  D(n) +1 -n,
\end{eqnarray}
respectively. From any characteristics of the considered model in the range \mbox{$0 \leq n \leq 1$} we can find the behaviour of the system in the range \mbox{$1 \leq n \leq 2$}. In particular, the phase diagrams are symmetric with respect to half-filling.


\section{Some analytical  VA results}\label{app:anal}

From the set \mbox{(\ref{row:MFA1})--(\ref{row:MFA2})} one can obtain the relation between the chemical potential and the electron concentration in both homogeneous phases:
\begin{eqnarray}
\label{row:miNO}\bar{\mu}_{NO} & =&  \frac{U}{2} + \\
&+ &\frac{1}{\beta}\ln \frac{n-1 + \sqrt{(n-1)^2 - n(n-2)\exp{(-\beta U)}}}{2-n},\nonumber\\
\label{row:miSS}\bar{\mu}_{SS} &= &I_0(n-1).
\end{eqnarray}

The formula determining the second order
\mbox{SS--NO} transition temperature $T_{SS}$ (the limit
\mbox{$\Delta\rightarrow0$} of equations (\ref{row:MFA1})--(\ref{row:MFA2})) has the following
form
\begin{equation}
\label{row:2ndorderTSSNO}
\beta_{SS} = \frac{2}{U} \ln \left[ \frac{\sinh\left(\beta_{SS}|\bar{\mu}_{SS}|\right)}{|n-1|} - \cosh\left(\beta_{SS} \bar{\mu}_{SS}\right)\right],
\end{equation}
where \mbox{$\beta_{SS}=1/(k_B T_{SS})$}.

Let us stress that $T_{SS}$ determined
from the above equation can be not only a~transition
temperature between stable phases but also a~transition
temperature between metastable phases (e.~g. in the regions
of PS states stability) or unstable phases. However,
all second order transition temperatures which were obtained
in section~\ref{sec:kT} by comparison of free energies
fulfill (\ref{row:2ndorderTSSNO}). From (\ref{row:2ndorderTSSNO}) one can obtain $T_{SS}$ as
a function of $n$ (or $\bar{\mu}$ -- cf. (\ref{row:miSS})). In the limiting cases it yields
\begin{eqnarray}
k_B T_{SS}(U\rightarrow-\infty)& = & 2k_BT_{SS}(U=0)= \nonumber \\
 & = & 2 I_0 (n-1)\ln^{-1}\left(\frac{n}{2-n}\right).\qquad
\end{eqnarray}

The TCP-line is described by the following equation:
\begin{equation}
\tanh(\beta_{\textrm{T}}I_0|n-1|) = \left(\frac{1}{|n-1|\beta_{\textrm{T}}I_0} + |n-1| \right)^{-1}.
\end{equation}
Projections of TCP points
on the $k_BT/I_0$-$\bar{\mu}/I_0$  plane are shown in
figure~\ref{rys:PSkTvsmi} by dashed-dotted curve.




\clearpage

\begin{thebibliography}{10}
\bibitem{MRR1990}
Micnas~R, Ranninger~J and Robaszkiewicz~S 1990 \textit{Rev. Mod. Phys.} \textbf{62} 113

\bibitem{AAS2010}
Johnston~D~C 2010 \textit{Adv. in Physics} \textbf{59} 803

\bibitem{PIN2009}
Park J T \etal 2009 \textit{Phys. Rev. Lett.} \textbf{102} 117006

\bibitem{RPC2011}
Ricci~A \etal 2011 \textit{Phys. Rev.}~B \textbf{84} 060511(R)

\bibitem{XTP2008}
Xu~J, Tan~S, Pi~L and Zhang~Y 2008 \textit{J. Appl. Phys.} \textbf{104} 063914

\bibitem{UAC2009}
Udby~L, Andersen~N~H, Chou~F~C, Christensen~N~B, Emery~S~B, Lefmann~K, Lynn~J~W, ~Mohottala~H~E, Niedermayer~C and Wells~B~O 2009 \textit{Phys. Rev.}~B \textbf{80} 014505

\bibitem{SFG2002}
Savici~A~T \etal 2002 \textit{Phys. Rev.}~B \textbf{66} 014524

\bibitem{MWB2006}
Mohottala~H~E, Wells~B~O, Budnick~J~I, Hines~W~A, Niedermayer~C, Udby~L, Bernhard~C, Moodenbaugh~A~R and Chou~F~C 2006 \textit{Nature Mat.} \textbf{5} 377

\bibitem{PNB2001}
Pan~S~H \etal 2001 \textit{Nature} \textbf{413} 282

\bibitem{KPK2004}
Kornilov~A~V, Pudalov~V~M, Kitaoka~Y, Ishida~K, Zheng~G-q, Mito~T and Qualls~J~S 2004 \textit{Phys. Rev.}~B \textbf{69}, 224404

\bibitem{CSP2008}
Colin~C~V, Salameh~B, and Pasquier~C~R 2008 \textit{J.~Phys.: Condens. Matter} \textbf{20} 434230

\bibitem{TCS2008}
Taylor~O~J, Carrington~A and Schlueter~J~A 2008 \textit{Phys. Rev.}~B \textbf{77} 060503(R)

\bibitem{MR1992}
Micnas~R and Robaszkiewicz~S 1992 \textit{Phys. Rev.}~B \textbf{45} 9900

\bibitem{MRK1995}
Micnas~R, Robaszkiewicz~S and Kostyrko~T 1995 \textit{Phys. Rev.}~B \textbf{52} 6863

\bibitem{BBM2002}
Bernardet~K, Batrouni~G~G, Meunier~J-L, Schmid~G, Troyer~M and Dorneich~A 2002
\textit{Phys. Rev.}~B \textbf{65} 104519

\bibitem{HD1993}
Hui~A and Doniach~S 1993 \textit{Phys. Rev.}~B \textbf{48} 2063

\bibitem{RB1999}
Robaszkiewicz~S and Bu\l{}ka~B~R 1999 \textit{Phys. Rev.}~B \textbf{59} 6430

\bibitem{JKS2001}
Japaridze~G~I, Kampf~A~P, Sekania~M, Kakashvili~P and Brune~Ph 2001 \textit{Phys. Rev.}~B \textbf{65} 014518
\item[]
Japaridze~G~I and M\"uller-Hartmann~E 1997 \textit{J.~Phys.: Condens. Matter} \textbf{9} 10509

\bibitem{CR2001}
Czart~W~R and Robaszkiewicz~S 2001 \textit{Phys. Rev.}~B \textbf{64} 104511
\item[]
Robaszkiewicz~S and Czart~W~R 2001 \textit{Acta Phys. Pol.}~B \textbf{32} 3267
\item[]
Robaszkiewicz~S and Czart~W~R 2003 \textit{Phys. Status Solidi}~B \textbf{236} 416
\item[]
Czart~W~R and Robaszkiewicz~S 2004 \textit{Acta Phys. Pol.}~A \textbf{106} 709

\bibitem{DM2000}
Dolcini~F and Montorsi~A 2000 \textit{Phys. Rev.}~B \textbf{62} 2315

\bibitem{Z2005}
Ziegler~K 2005 \textit{Laser Physics} \textbf{15} 650

\bibitem{MM2004}
Mierzejewski~M and Ma\'ska~M~M 2004 \textit{Phys. Rev.}~B \textbf{69} 054502

\bibitem{MC1996}
Montorsi~A and Campbell~D~K 1996 \textit{Phys. Rev.}~B \textbf{53} 5153

\bibitem{H1963}
Hubbard~J 1963 \textit{Proc. R. Soc. London} Ser.~A \textbf{276} 238
\item[]
Campbell~D~K, Gammel~J~T and Loh~E~Y 1990 \textit{Phys. Rev.}~B \textbf{42} 475
\item[]
Hirsch~J~E 1991 \textit{Physica}~C \textbf{179} 317
\item[]
Amadon~J~C and Hirsch~J~E 1996 \textit{Phys. Rev.}~B \textbf{54} 6364

\bibitem{FH1983}
Fradkin~E and Hirsch~J~E 1983 \textit{Phys. Rev.}~B \textbf{27} 1680
\item[]
Miyake~K, Matsuura~T, Jichu~H and Nagaoka~Y 1983 \textit{Prog. Theor.  Phys.} \textbf{72} 1063

\bibitem{RMR1987}
Robaszkiewicz~S, Micnas~R and Ranninger~J 1987 \textit{Phys. Rev.}~B \textrm{36} 180

\bibitem{BL1988}
Bastide~C and Lacroix~C 1988 \textit{J.~Phys.}~C \textrm{21} 3557

\bibitem{B1973}
Bari~R~A 1973 \textit{Phys. Rev.}~B \textbf{7} 2128

\bibitem{HB1977}
Ho W-C and Barry~J~H 1977 \textit{Phys. Rev.}~B \textbf{16} 3172

\bibitem{WA1987}
Wiecko~C and Allub~R 1987 \textit{Phys. Rev.}~B \textbf{35} 2041

\bibitem{RP1993}
Robaszkiewicz~S and Paw\l{}owski~G 1993 \textit{Physica}~C \textbf{210} 61

\bibitem{R1994}
Robaszkiewicz~S 1994 \textit{Acta Phys. Pol.}~A \textbf{85} 117

\bibitem{PTVF1992}
Press~W~H, Teukolsky~S~A, Vetterling~W~T  and Flannery~B~P 1992 \emph{Numerical Recipes
in C. The Art of Scientific Computing}, Second Edition, (Cambridge: Cambridge University Press)

\bibitem{AS1991}
Arrigoni~E and Strinati~G~C 1991 \textit{Phys. Rev.}~B \textbf{44} 7455

\bibitem{B2004}
B\k{a}k~M 2004 \textit{Acta Phys. Pol.}~A \textbf{106} 637

\bibitem{LSM1961}
Lieb~E, Schultz~T and Mattis~D 1961 \textit{Ann. Phys.} \textbf{16} 407

\bibitem{K1962}
Katsura~S 1962 \textit{Phys. Rev.} \textbf{127} 1508

\bibitem{N1967}
Niemeijer~Th 1967 \textit{Physica} \textbf{36} 377

\bibitem{TOK1976}
Tsuchida~Y, Oguchi~A and Kanai~K 1976 \textit{Prog. Theor. Phys.} \textbf{56} 1011

\bibitem{K1966}
Kleiner~W~H 1966 \textit{Phys. Rev.} \textbf{142} 318

\bibitem{SH1973}
Semura~J~S and Huber~D~L 1973 \textit{Phys. Rev.}~B \textbf{7} 2154

\bibitem{M0000}
Kapcia~K, Micnas~R and Robaszkiewicz~S, in preparation

\bibitem{MW1966}
Mermin~N~D and Wagner~H 1966 \textit{Phys. Rev. Lett.} \textbf{17} 1133

\bibitem{RP1996}
Robaszkiewicz~S and Paw\l{}owski~G 1996 \textit{Acta. Phys. Pol.}~A \textbf{90} 569

\bibitem{JS2002}
Japaridze~G~I and Sarkar~S 2002 \textit{Eur. Phys. J.}~B. \textbf{27} 139

\bibitem{MM2008}
Mancini~F and Mancini~F~P 2008 \textit{Phys. Rev.}~E \textbf{77} 061120
\item[]
Mancini~F and Mancini~F~P 2009 \textit{Eur. Phys. J.}~B \textbf{68} 341

\bibitem{EK1993}
Emery~V~J and Kivelson~S~A 1993 \textit{Physica}~C \textbf{209} 597

\bibitem{JCC2001}
Lorenzana~J, Castellani~C, and Di Castro~C 2001 \textit{Phys. Rev.}~B \textbf{64} 235127
\item[]
Lorenzana~J, Castellani~C, and Di Castro~C 2001 \textit{Phys. Rev.}~B \textbf{64} 235128

\bibitem{YY2004}
Yukalov~V~I and Yukalova~E~P 2004 \textit{Phys. Rev.}~B \textbf{70} 224516

\bibitem{RMC1984}
Robaszkiewicz~S, Micnas~R and Chao K~A 1984 \textit{Phys. Rev.}~B \textbf{29} 2784
\item[]
Kapcia~K, K\l{}obus~W and Robaszkiewicz~S 2010 \textit{Acta. Phys. Pol.}~A \textbf{118} 350
\item[]
Kapcia~K and Robaszkiewicz~S 2011 \textit{J.~Phys.: Condens. Matter} \textbf{23} 105601
\item[]
Kapcia~K and Robaszkiewicz~S 2011 \textit{J.~Phys.: Condens. Matter} \textbf{23} 249802

\bibitem{KKR2010}
K\l{}obus~W, Kapcia~K and Robaszkiewicz~S 2010 \textit{ Acta Phys. Pol.}~A \textbf{118} 353
\item[]
Kapcia~K 2012 \textit{ Acta Phys. Pol.}~A \textbf{121} 733
\end{thebibliography}
\end{document}